\documentclass[journal]{IEEEtran}

\usepackage{graphicx}
\usepackage{multirow}
\usepackage{amsmath,amssymb}
\usepackage{url}
\usepackage{pdfpages}
\usepackage{latexml}
\usepackage{placeins}

\begin{document}

\title{The Lift Spectrum: How Measurement-to-Space Adaptivity Shapes Robustness in Image-Free Single-Pixel Sensing}

\author{Yuyuan Han, Jingwei Li, Xiaoxia Zhang, Long Qiu, Chong Wang, Wenxuan Hao, Jiangyu Han, Xinyu Yao, Yuchen He, Hui Chen, Jianbin Liu, and Huaibin~Zheng%
\thanks{Manuscript received XX XX, 2026; revised XX XX, 2026. (\textit{Corresponding author: Huaibin Zheng.})}%
\thanks{Yuyuan Han, Long Qiu, Chong Wang, Wenxuan Hao, Jiangyu Han, Xinyu Yao, Hui Chen, and Jianbin Liu are with the Electronic Materials Research Laboratory, Key Laboratory of the Ministry of Education and International Center for Dielectric Research, School of Electronic Science and Engineering, Xi'an Jiaotong University, Xi'an 710049, China.}%
\thanks{Yuchen He and Huaibin Zheng are with the Electronic Materials Research Laboratory, Key Laboratory of the Ministry of Education and International Center for Dielectric Research, School of Electronic Science and Engineering, Xi'an Jiaotong University, Xi'an 710049, China, and also with the National Key Laboratory of Human-Machine Hybrid Augmented Intelligence, National Engineering Research Center for Visual Information and Applications, and Institute of Artificial Intelligence and Robotics, Xi'an Jiaotong University, Xi'an 710049, China (e-mail: huaibinzheng@xjtu.edu.cn).}%
\thanks{Jingwei Li is with the Micius Laboratory, Henan Academy of Sciences, Zhengzhou 450052, China.}%
\thanks{Xiaoxia Zhang is with the Department of Pharmacy, the First Affiliated Hospital of Xi'an Jiaotong University, Xi'an 710061, China.}%
\thanks{This work has been submitted to the IEEE for possible publication. Copyright may be transferred without notice, after which this version may no longer be accessible.}}

\markboth{IEEE Transactions on Computational Imaging,~Vol.~XX, No.~X, Month~2026}%
{Han \MakeLowercase{\textit{et al.}}: The Lift Spectrum for Image-Free Single-Pixel Sensing}

\maketitle
\bstctlcite{BSTcontrol}

\begin{abstract}
Single-pixel sensing encodes a scene as a short sequence of coded measurements, and image-free methods infer the task directly from that sequence. We show that removing image reconstruction relocates the central design problem to the \emph{lift}: how 1D measurements become a 2D task representation. We organize this choice as a \emph{lift spectrum} from a fixed-physics inverse, through a learned static projection, to content-adaptive retrieval. These are not interchangeable forms of reconstruction: the fixed-physics route reconstructs an image consumed at inference, whereas our spatiotemporal soft-fusion (STSF) network lifts measurements directly into task features, and task-prioritized loss scheduling (TPLS) uses a separate learned reconstruction branch only as scheduled training supervision. A probe-selected recurrent encoder and a parameter-matched lift ablation identify the STSF design. In simulation, STSF+TPLS exceeds the prior image-free baseline on three datasets at 3.13\% sampling (+3.2 to +9.9~pp foreground mIoU) and remains competitive down to 0.39\%. The strongest clean-trained reconstruct-then-segment baseline wins without measurement noise, but measurement noise reverses the ranking: the reconstructed task input carries a 20--70$\times$ larger normalized relative perturbation than the measurements themselves. Stressed to failure, the three lift regions exhibit distinct dominant signatures---collapse, imprinting, and coarsening. STSF+TPLS transfers without fine-tuning to a real single-pixel bench, where the reversal reappears as a proof of concept; inference takes about 14~ms per mask on an RTX 4090. Within the tested fixed-acquisition regime, measurement-to-space adaptivity therefore organizes both the clean-to-noisy operating envelope and the failure a system encounters. Code and pretrained weights: \url{https://github.com/Hanyuyuan6/STSF-TPLS}.
\end{abstract}


\section{Introduction}
\label{sec:intro}

\IEEEPARstart{V}{ision} at the edge, on low-power sensor nodes and the agents that perceive through them, is often limited less by computation than by data movement. Acquiring, storing, and transmitting dense images dominates the energy budget of a low-power sensor, even though most tasks use only a small fraction of what an image contains. Single-pixel sensing (SPS), with roots in ghost imaging \cite{bc1}, removes this overhead at its source by compressing the measurement in hardware. A sequence of spatially modulated patterns illuminates the scene, and a single bucket detector records one intensity per pattern, encoding a 2D scene as a 1D measurement sequence \cite{bc2,tci1}. Concentrating the scene's light onto a single detector permits simple, inexpensive detectors \cite{bc6}, tolerates scattering media \cite{bc3}, admits noise-hardened measurement-side pattern design \cite{bc4}, pairs with well-conditioned fast-transform Hadamard or Fourier bases \cite{alg1,alg3}, and extends to beyond-optical modalities \cite{bc7} and to spectral bands where detector arrays are costly \cite{bc8}. It suits continuous perception on bandwidth- and power-constrained platforms \cite{if1,if12}: at the 3.13\% sampling rate used throughout, the sensor emits 512 coded measurements per decision instead of $128{\times}128$ pixels, a $32\times$ reduction in acquired samples.

Most deep learning for SPS reconstructs the scene image from its measurements \cite{review1}. For many edge tasks the image is only an intermediate; what is needed is the segmentation mask, the detection, or the class. Inferring the task directly from the 1D measurements, without forming an image, removes the most expensive stage of the pipeline \cite{ifbc1,ifbc2,if3}. This image-free route, demonstrated from classification through tracking to dense segmentation (Sec.~\ref{sec:related}), establishes that task-specific information survives in the compressed measurements. Throughout, image-free refers to inference: at test time the network maps the measurement sequence straight to the task output.

Removing reconstruction relocates the difficulty to the \emph{lift}, the step that turns a 1D measurement sequence into a 2D representation the task network can read. This step is unavoidable: a bucket detector is spatially zero-dimensional, so every image-free system must decide how 1D evidence becomes 2D space. The closest image-free segmentation baseline \cite{SOTA} makes that decision with a static dense projection. Such a lift applies the same weights to every sequence and cannot retrieve measurements according to the scene at hand. A capacity-controlled comparison shows that gated recurrent encoders preserve more scene structure than a matched dense map, while a parameter-matched comparison of six retrieval modules isolates content-adaptive cross-attention. A second difficulty is optimization: direct task supervision from extreme compression is weakly constrained, whereas a static auxiliary-reconstruction weight can under-use that prior early and over-regularize the task head late.

Both difficulties meet at the same under-examined interface. We therefore chart the lift along two axes---how the measurement sequence is \emph{encoded} and how it is \emph{retrieved} into space. The selected recurrent encoder and content-adaptive cross-attention retrieval define the spatiotemporal soft-fusion (STSF) network, whose compact lift feeds a U-Net++ decoder \cite{unet++} (Sec.~\ref{subsec:stsf}). This learned measurement-to-space lift is not a learned image reconstructor: the lifted features go directly to the task decoder and are never rendered as an inference image.

Reconstruction re-enters only off the mask-prediction path. It serves first as the task-agnostic probe used to select the temporal encoder and then as a scheduled training prior: task-prioritized loss scheduling (TPLS) applies a reconstruction loss through a parallel auxiliary head, shaping the shared representation early before yielding priority to segmentation \cite{physml}. In the tested extreme-compression setting, the reconstructed output is therefore not the deployed task input; reconstruction is structured supervision for learning the direct lift. Soft fusion names this scheduled blending of acquisition prior and task objective.

We test this account at 3.13\% sampling, across a $32\times$ span down to $0.39\%$, and on a real single-pixel optical bench. The resulting lift spectrum compares a fixed-physics reconstruct-then-segment route, a learned static projection, and STSF's content-adaptive retrieval. Task-adapted reconstruction leads without measurement noise; under the tested measurement-noise calibration, image-free decoding overtakes it because the reconstructed input delivers a 20--70$\times$ larger normalized relative perturbation to its task head at 20~dB. The bench echoes this reversal as a proof of concept. As acquisition degrades, the three designs fail in characteristically different ways---collapse, imprinting, and coarsening (Sec.~\ref{subsec:failure}). Our central claim is therefore bounded but stronger than a method comparison: within the tested fixed-acquisition regime, measurement-to-space adaptivity organizes the clean-to-noisy operating envelope.

\section{Related Work}
\label{sec:related}

\noindent\looseness=-1\textbf{Learned reconstruction for SPS.} Deep single-pixel imaging primarily maps compressive measurements back to an image. The field has progressed from post-reconstruction enhancement \cite{GIDL2,GIDL3} and direct inversion with learned patterns \cite{GIDL4} or simulation-trained inverse mappings \cite{GIDL5} to recurrent sequence models \cite{GIDL6RNN}, physics-enhanced objectives \cite{GIDL8PE1}, unrolled solvers \cite{tci2,proxunroll}, generative priors \cite{GIDL10PE2}, transformers \cite{GIDL11transformer}, and differentiable dynamics for changing scenes \cite{tci3}. These methods improve the image as an inference product. Our use of learned reconstruction is categorically different: it is a probe for encoder selection (Sec.~\ref{subsec:benchmark}) and an auxiliary objective scheduled by TPLS, never the image consumed by the segmentation head.

\noindent\looseness=-1\textbf{Image-free SPS.} Direct inference from compressive measurements has progressed from action and class recognition \cite{ifbc1,ifbc2} and ghost cytometry \cite{ifbc3} to classification \cite{if5}, detection and tracking \cite{if8,if9,if12}, saliency and invariant moments \cite{if7,if11}, dense segmentation \cite{SOTA,ifseg2}, and pose estimation \cite{ifpose}; complementary work improves sampling or label efficiency \cite{if1,if2,if3}. In the dense image-free systems closest to ours, however, the measurement-to-space step remains input-independent: SPIFS uses a dense projection of the flattened measurements, and the multi-scale U-Net variant uses a shallow learned lift. Dynamic-scene tracking instead reads Fourier measurements through classical filtering and optical flow \cite{iftrack}. This work makes the lift itself the controlled variable, tests content-adaptive retrieval, and asks how lift choice changes the operating regime.

\noindent\looseness=-1\textbf{Reconstruct-then-task and task adaptation.} The classical alternative reconstructs an image before applying a task network \cite{dl_imaging}. Direct prediction from undersampled data \cite{joint_recon_seg_bypass}, joint reconstruction--segmentation \cite{joint_csmri_seg}, task-aware compressed sensing \cite{task_aware_cs}, and task-driven restoration \cite{task_driven_restore} all question whether image fidelity is the right intermediate objective. This concern is reinforced by evidence that learned reconstructors can be unstable \cite{recon_instab} or generate plausible false structure \cite{recon_halluc}, raising risks not captured by image-fidelity metrics alone. Our task-adapted (TA) baseline gives the two-stage route its strongest controlled form by retraining the task head on the reconstructions it will consume. It wins under clean acquisition, then loses under tested measurement noise when that reconstructed-input distribution shifts (Sec.~\ref{subsec:regime}). The contribution is thus an operating map, not a claim that one route dominates universally.

\noindent\looseness=-1\textbf{Physical priors, scheduling, and learnable queries.} Physical knowledge can enter through data, architecture, or the learning objective \cite{physml}. Single-pixel inversion commonly embeds the forward operator through unrolling \cite{tci2} or physics-enhanced losses \cite{GIDL8PE1}. TPLS instead supplies a structured auxiliary reconstruction objective and anneals its influence, a discrete continuation strategy \cite{continuation} that also moderates interference between reconstruction and segmentation gradients \cite{ad1}. The STSF lift adapts the learnable-query decoding of Perceiver and DETR \cite{perceiver,detr} to measurement-to-space retrieval. Finally, joint source--channel coding \cite{deep_jscc} and task-oriented communication \cite{task_oriented_comm} provide a useful analogy: end-to-end task objectives can degrade more gracefully than pipelines that must first recover an intermediate signal. We test that analogy here rather than assume it.
\section{Methods}

\subsection{Spatiotemporal Soft-Fusion Network}
\label{subsec:stsf}

\begin{figure*}[t]
\centering\includegraphics[width=\linewidth]{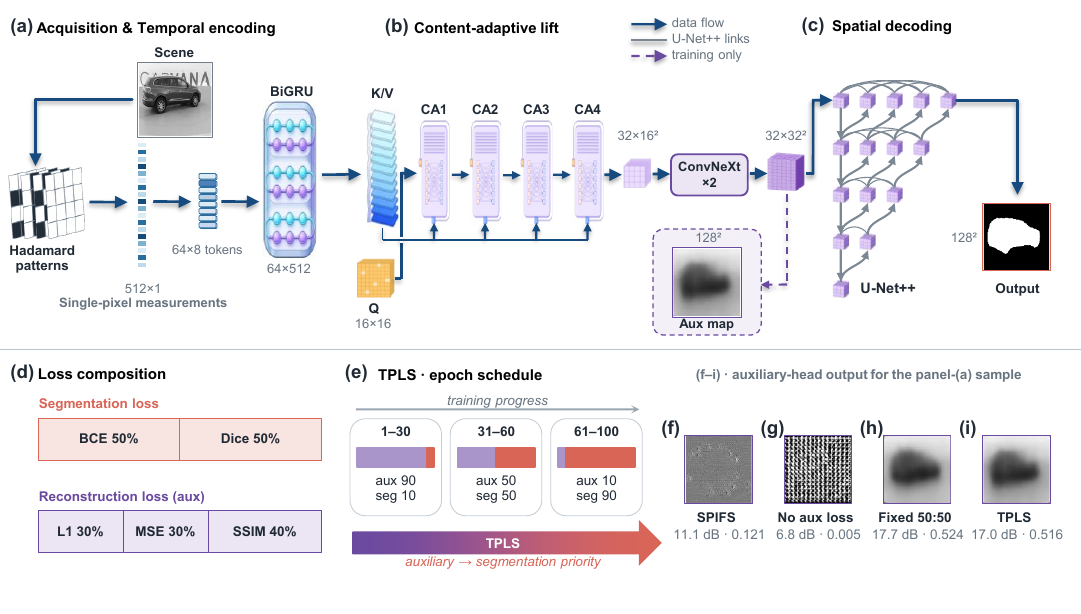}
\caption{Proposed STSF+TPLS architecture for image-free segmentation from a 1D single-pixel measurement sequence.
(a) Acquisition and temporal encoding: natural-order Sylvester Hadamard patterns ($\pm1$) give 512 measurements at a 3.13\% sampling rate (min--max normalized per sample), which are tokenized, linearly projected, and encoded by a three-layer bidirectional GRU into key/value tokens.
(b) Content-adaptive lift: a learned $16{\times}16$ spatial query grid ($\mathrm{Q}$) attends to these tokens through four cross-attention blocks (CA1--CA4; 8 heads, $d{=}288$; Eqs.~\eqref{eq:lift_block}--\eqref{eq:lift_attn}), producing a compact spatial feature.
(c) Spatial decoding: ConvNeXt blocks and a U-Net++ decoder expand the shared feature to the $1{\times}128^2$ foreground mask. A parallel auxiliary head reconstructs $X_{\mathrm{afm}}$ from the same shared feature; its scheduled loss shapes that feature during training, but $X_{\mathrm{afm}}$ is not consumed by the segmentation head. The released forward pass retains and returns this prunable auxiliary output.
(d) Loss composition, drawn to its exact component weights (Sec.~\ref{subsec:tpls}).
(e) TPLS priority schedule of Eq.~\eqref{eq:tpls_schedule}: phase boundaries are fractions of the epoch budget, instantiated here for Carvana.
(f--i) Auxiliary-head output for the panel-(a) Carvana sample at each configuration's best-validation checkpoint (PSNR/SSIM against that input): (f) SPIFS; (g) no auxiliary loss; (h) fixed $50{:}50$; (i) TPLS. Two regimes appear, not four: without a weighted auxiliary loss the head does not reconstruct (f,~g), any nonzero weight recovers the object (h,~i; Sec.~\ref{subsec:sim}). These are diagnostic probes, not predictions.}
\label{fig:STSF}
\end{figure*}

STSF rests on one observation: under extreme compression the bottleneck is not the spatial decoder, which mature dense-prediction architectures handle, but the mapping from a 1D measurement sequence to a spatially organized, task-relevant representation. The lift spans the two axes above, encoding and retrieval; prior image-free methods sit at one corner of this space (a generic encoder and a dense reshape) and STSF at the opposite, each axis fixed by a dedicated study: the encoder by the reconstruction probe of Sec.~\ref{subsec:benchmark}, the retrieval by the parameter-matched ablation of Sec.~\ref{subsec:sim}.

We model SPS acquisition with a standard linear forward model,
\begin{equation}
s = \Phi x + n,
\label{eq:sps_forward}
\end{equation}
where $x=\mathrm{vec}(X_{\mathrm{img}})\in\mathbb{R}^{HW}$ is the vectorized scene, $s\in\mathbb{R}^{M}$ is the measurement sequence, $n\in\mathbb{R}^{M}$ is the measurement noise, and $\Phi\in\mathbb{R}^{M\times HW}$ is the sensing matrix whose rows correspond to the vectorized modulation patterns used in acquisition. In this work, the pattern schedule follows the natural-order (Sylvester) Hadamard construction, fixed and shared across all compared methods.

\noindent\textbf{Information equivalence under the shared acquisition.} Every compared method observes the \emph{same} acquisition of Eq.~\eqref{eq:sps_forward} for a given scene---identical patterns, ordering, and per-sample normalization---so the reported performance differences cannot stem from differential access to the scene: the design lever is the lift, not the sensing. Information equivalence does not imply equal achievable accuracy: a better lift extracts more task structure from the same $s$.

Given this measurement sequence $s$, our goal is to infer a segmentation mask $\hat{X}_{\mathrm{mask}}\in[0,1]^{H\times W}$ directly from the 1D input, without reconstructing an intermediate image for the segmentation head. To facilitate stable learning, STSF also outputs an image-like auxiliary map $X_{\mathrm{afm}}$ (Fig.~\ref{fig:STSF}(c)). Both heads read the same lifted feature: the reconstruction loss therefore updates the shared representation during training, whereas $X_{\mathrm{afm}}$ itself is never fed to the segmentation head. The released implementation still computes and returns this output; the auxiliary head is prunable for deployment.

The network follows the three stages of Fig.~\ref{fig:STSF}: temporal encoding (a), a content-adaptive lift (b), and spatial decoding (c). The temporal encoder is a standard gated recurrent unit (GRU)~\cite{gru}, selected by the capacity-controlled comparison of eight architectures in Sec.~\ref{subsec:benchmark}, in which a plain dense baseline at a matched parameter budget is consistently outperformed.

We then lift the temporally encoded 1D representation into a spatial feature map suitable for dense prediction. The length-$M$ input is partitioned into $T$ equal-length tokens of length $P$, where $M=T\cdot P$ (here $T{=}64$ and $P{=}8$ for $M{=}512$). Each token is linearly projected into a latent space (layer normalization, GELU, dropout) and a three-layer bidirectional GRU with hidden width $256$ encodes the token sequence. Rather than pooling the recurrent states into a single descriptor and broadcasting it through a fixed dense projection, we cast the 1D-to-2D lift as a \emph{content-adaptive cross-attention} retrieval with learnable queries~\cite{perceiver,detr}. The GRU token matrix $G=[h_1,\dots,h_T]$ is linearly projected and given a learned temporal position embedding, $\tilde{H}=G\,W_{kv}+E_{\mathrm{pos}}\in\mathbb{R}^{T\times d}$ ($d{=}288$), where $W_{kv}$ is a shared key/value pre-projection that the per-head $W_k^i,W_v^i$ of Eq.~\eqref{eq:lift_attn} further specialize, and $E_{\mathrm{pos}}$ encodes the acquisition order explicitly. A learnable grid of $N_{q}{=}16{\times}16$ spatial queries $Q^{(0)}\in\mathbb{R}^{N_{q}\times d}$ then reads from these tokens through $L{=}4$ pre-norm transformer blocks, each a cross-attention sublayer followed by a feed-forward sublayer,
\begin{equation}
\begin{aligned}
\tilde{Q}^{(l)} &= Q^{(l-1)} + \mathrm{MHCA}\!\bigl(\mathrm{LN}(Q^{(l-1)}),\,\mathrm{LN}(\tilde{H})\bigr),\\
Q^{(l)} &= \tilde{Q}^{(l)} + \mathrm{FFN}\!\bigl(\mathrm{LN}(\tilde{Q}^{(l)})\bigr),\quad l=1,\dots,L,
\end{aligned}
\label{eq:lift_block}
\end{equation}
where $\mathrm{LN}$ is layer normalization, $\mathrm{FFN}$ is a two-layer GELU network of hidden width $4d$, and the $i$-th of the eight attention heads in $\mathrm{MHCA}$, each of per-head dimension $d_h{=}d/8{=}36$, is
\begin{equation}
\mathrm{head}_i = \mathrm{softmax}\!\left(\frac{(Q W_q^{i})(\tilde{H} W_k^{i})^{\!\top}}{\sqrt{d_h}}\right)\tilde{H} W_v^{i},
\label{eq:lift_attn}
\end{equation}
with learnable projections $W_q^{i},W_k^{i},W_v^{i}$; here $Q$ and $\tilde{H}$ are the layer-normalized inputs $\mathrm{LN}(Q^{(l-1)})$ and $\mathrm{LN}(\tilde{H})$ of Eq.~\eqref{eq:lift_block}. The output queries $Q^{(L)}$ are layer-normalized, reshaped onto the $16{\times}16$ grid, and projected by a $1{\times}1$ convolution to a $C{\times}16{\times}16$ feature map ($C{=}32$); relying on attention rather than a large dense flattening layer keeps the parameter count compact. The feature map is then refined by two lightweight ConvNeXt blocks~\cite{convnext}, which add local mixing and optimization stability, adaptively upsampled, and decoded by a U-Net++ decoder~\cite{unet++}; the network emits pixel-wise logits, and a sigmoid followed by a $p>0.5$ threshold yields the final mask $\hat{X}_{\mathrm{mask}}$ at evaluation.

The cross-attention retrieval makes the 1D-to-2D assignment content-dependent through the softmax over $\tilde{H}$, with $E_{\mathrm{pos}}$ tagging each token by its acquisition index; under extreme compression this yields higher downstream accuracy than a fixed projection (Sec.~\ref{subsec:sim}).

In the network's name, \emph{spatiotemporal} describes the representation---recurrent integration along the acquisition axis, spatial organization after the lift---and \emph{soft fusion} the adaptive training-time prioritization between the auxiliary reconstruction and segmentation objectives (Sec.~\ref{subsec:tpls}).

\subsection{Physics-Guided Task-Prioritized Loss Scheduling}
\label{subsec:tpls}

Fixed loss weights cannot prioritize different objectives at different stages of optimization: a fixed weight may over-regularize the task head when the auxiliary term dominates, or leave the auxiliary branch undertrained, consistent with gradient-interference effects in multi-task learning~\cite{ad1}. The fixed-weight auxiliary-loss (fixed-AL) ablation (Sec.~\ref{subsec:sim}) isolates the value of \emph{scheduling} over a fixed weight, and a direct measurement along the training trajectory supports the \emph{magnitude} half of this reading: the two gradients stay near-orthogonal throughout (mean $\cos(\nabla\mathcal{L}_{\mathrm{Seg}},\nabla\mathcal{L}_{\mathrm{AL}}){\approx}0$, so hard directional conflict is rare), but their magnitude ratio $\lVert\nabla\mathcal{L}_{\mathrm{AL}}\rVert/\lVert\nabla\mathcal{L}_{\mathrm{Seg}}\rVert$ rises from ${\sim}0.3$ early to ${\sim}1.7$ late as the task loss converges, crossing unity mid-training. A \emph{fixed} weight therefore lets the auxiliary term dominate precisely when the task should lead; the TPLS schedule keeps the effective pull $\alpha\lVert\nabla\mathcal{L}_{\mathrm{AL}}\rVert/\beta\lVert\nabla\mathcal{L}_{\mathrm{Seg}}\rVert$ bounded and ultimately decreasing (${\approx}2.4\!\to\!1.0\!\to\!0.2$; means over five repeats, protocol and per-phase ranges in Supplementary Sec.~S6).

We accordingly propose physics-guided task-prioritized loss scheduling (TPLS), a stage-wise schedule that reweights the auxiliary and task losses: reconstruction-oriented supervision dominates early, establishing acquisition-consistent priors in the shared representation, then yields to the task as discriminative structure emerges (the continuation-scheme reading of Sec.~\ref{sec:related}). In the taxonomy of Sec.~\ref{sec:related}, this physics guidance is observational---a learned measurement-to-scene inversion supervised by measurement--scene pairs---modulated by the schedule as a learning bias, not a forward-operator penalty as in physics-informed neural networks. The resulting loss-specific and validation dynamics are reported in Supplementary Fig.~S2.

We define the auxiliary loss (AL) as an image-domain composite objective that encourages both pixel-wise fidelity and structural consistency between the auxiliary map $X_{\mathrm{afm}}$ and the ground-truth image $X_{\mathrm{img}}$. As established in Sec.~\ref{subsec:stsf}, supervision from this branch is used only during training:
\begin{equation}
\label{eq:AL}
\begin{aligned}
\mathcal{L}_{\mathrm{AL}}(X_{\mathrm{afm}},X_{\mathrm{img}})
&= \lambda_{1}\,\mathcal{L}_{1}(X_{\mathrm{afm}},X_{\mathrm{img}}) \\
&\quad + \lambda_{2}\,\mathcal{L}_{2}(X_{\mathrm{afm}},X_{\mathrm{img}}) \\
&\quad + \lambda_{\mathrm{ssim}}\,\mathcal{L}_{\mathrm{ssim}}(X_{\mathrm{afm}},X_{\mathrm{img}}).
\end{aligned}
\end{equation}
We set $\lambda_{1}=\lambda_{2}=0.3$ and $\lambda_{\mathrm{ssim}}=0.4$. Here $\mathcal{L}_{1}$ and $\mathcal{L}_{2}$ are the mean absolute error and the mean squared error between $X_{\mathrm{afm}}$ and $X_{\mathrm{img}}$ (the auxiliary map resized to match when needed), and $\mathcal{L}_{\mathrm{ssim}}=\tfrac{1}{2}\bigl(1-\mathrm{SSIM}(X_{\mathrm{afm}},X_{\mathrm{img}})\bigr)$, the structural-dissimilarity (DSSIM) form, with SSIM computed over $5{\times}5$ Gaussian windows.

The total training loss at epoch $e$ is
\begin{equation}
\begin{aligned}
\mathcal{L}_{\mathrm{total}}(e)
&= \alpha(e)\,\mathcal{L}_{\mathrm{AL}}\!\big(X_{\mathrm{afm}},X_{\mathrm{img}}\big) \\
&\quad + \beta(e)\,\mathcal{L}_{\mathrm{Seg}}\!\big(\hat{X}_{\mathrm{mask}},X_{\mathrm{mask}}\big),
\end{aligned}
\label{eq:total_loss}
\end{equation}
where $X_{\mathrm{afm}}$ denotes the auxiliary map (Fig.~\ref{fig:STSF}(c)), $X_{\mathrm{img}}$ the ground-truth image, $\hat{X}_{\mathrm{mask}}$ the predicted mask, $X_{\mathrm{mask}}$ the ground-truth mask, and $\alpha(e),\beta(e)$ are time-varying loss weights specified by TPLS (Eq.~\eqref{eq:tpls_schedule}).

Because $\mathcal{L}_{\mathrm{total}}(e)$ is linear in its two terms, its parameter gradient is the convex combination $\alpha(e)\,\nabla_{\Theta}\mathcal{L}_{\mathrm{AL}}+\beta(e)\,\nabla_{\Theta}\mathcal{L}_{\mathrm{Seg}}$ ($\alpha+\beta=1$), so scheduling $(\alpha,\beta)$ shifts the dominant optimization signal from $\mathcal{L}_{\mathrm{AL}}$ to $\mathcal{L}_{\mathrm{Seg}}$ across training.

The segmentation loss $\mathcal{L}_{\mathrm{Seg}}$ combines the standard Dice and binary cross-entropy (BCE) losses with equal weights $w_{\mathrm{Dice}}=w_{\mathrm{BCE}}=0.5$, computed over all predicted pixels in the batch; Dice emphasizes region overlap and tolerates foreground--background imbalance, while BCE calibrates per-pixel probabilities. For the three-class white-blood-cell (WBC) data, the sigmoid and BCE are replaced by a softmax and multi-class cross-entropy, with the Dice term averaged over classes.

TPLS specifies a schedule over the loss weights $(\alpha(e),\beta(e))$ across training. Let the total number of epochs be $E$. We define phase boundaries $e_{1}=\lfloor r_{1}E\rfloor$ and $e_{2}=\lfloor (r_{1}+r_{2})E\rfloor$ with phase fractions $(r_{1},r_{2})$ specified per dataset in Supplementary Sec.~S1. We keep $(\alpha(e),\beta(e))$ fixed within each phase:
\begin{equation}
(\alpha(e),\beta(e))=
\begin{cases}
(0.9,\,0.1), & 1\le e \le e_{1},\\
(0.5,\,0.5), & e_{1}< e \le e_{2},\\
(0.1,\,0.9), & e_{2}< e \le E,
\end{cases}
\label{eq:tpls_schedule}
\end{equation}
where $e$ denotes the current epoch and $\lfloor\cdot\rfloor$ is the floor operator.

\section{Experiments}

The evidence follows the paper's central design question: a shared acquisition and evaluation protocol (Sec.~\ref{sec:setup}); the two lift axes, supported first by temporal-encoder selection and then by end-to-end retrieval and component ablations (Secs.~\ref{subsec:benchmark}--\ref{subsec:sim}); the clean-to-noisy operating regime (Sec.~\ref{subsec:regime}); the resulting failure signatures (Sec.~\ref{subsec:failure}); and transfer to a real single-pixel bench (Sec.~\ref{subsec:phy}).


\subsection{Experimental Setup}
\label{sec:setup}

We evaluate image-free SPS segmentation on three datasets: the Carvana vehicle-masking benchmark \cite{dataset1}, MNIST handwritten digits \cite{dataset2}, and a WBC microscopy dataset \cite{dataset5}. The temporal-encoder probe of Sec.~\ref{subsec:benchmark} (Table~\ref{tab:probe}) instead uses MNIST, Fashion-MNIST, and CelebA, standard benchmarks on which reconstruction fidelity is assessed.

\noindent\textbf{Datasets and preprocessing.} The three datasets cover complementary segmentation regimes: Carvana provides binary vehicle masks; MNIST is cast as binary digit-foreground segmentation; and the WBC dataset provides three-class masks (background, cytoplasm, and nucleus). MNIST segmentation is trivial in the image domain, where a threshold suffices, but not from 512 compressed measurements where no image exists: the off-the-shelf reconstruct-then-segment baseline (Supplementary Fig.~S4) instantiates this image-domain route on the reconstructions and collapses, and MNIST's sparse strokes probe the regime of least redundancy, where a content-adaptive lift helps most. To avoid train/test leakage, the Carvana split is group-disjoint by vehicle identity and the MNIST and WBC splits are sample-disjoint, and all metrics are reported on held-out test splits. All images and masks are resized to $128{\times}128$ pixels, and the same preprocessing is applied to all methods; training-time augmentation is specified in Supplementary Sec.~S1, and validation/test images are unaugmented. Each measurement sequence is min--max normalized to $[0,1]$ per sample before entering the network, in simulation and on hardware alike.

\noindent\textbf{Acquisition model and sampling rate.} We simulate SPS acquisition using natural-order Hadamard patterns (Sylvester construction). Buckets are synthesized noiselessly and normalized per sample by min--max; this holds for the segmentation experiments and for the reconstruction probe of Table~\ref{tab:probe} alike. Measurement noise enters only at evaluation time, in the study of Sec.~\ref{subsec:regime}. This fixed ordering is shared by all methods; it is not sequency-ordered and is therefore not optimized for low-rate image reconstruction. We deliberately hold the acquisition fixed and unoptimized so that every difference between the compared methods is attributable to the lift. The sampling rate is defined as $M/(H\cdot W)$; for $H=W=128$, a 3.13\% rate corresponds to $M{=}512{=}2^{9}$ measurements, a natural power-of-two Hadamard subset sitting in the extreme-compression regime where classical reconstruction under this ordering is severely degraded as an image yet retains task-usable structure (Sec.~\ref{subsec:sim}), precisely the regime image-free sensing targets.

\noindent\textbf{Compared methods.} We evaluate two categories of baselines. (i) End-to-end image-free segmentation: SPIFS \cite{SOTA} and STSF variants (no AL, a fixed auxiliary weight of $0.5$, and TPLS). (ii) Reconstruct-then-segment: a compressed-sensing (CS) reconstruction \cite{bc2} solved via the alternating direction method of multipliers (ADMM) and Hadamard single-pixel imaging (HSI) \cite{alg1}, each followed by the same U-Net++ segmenter used in STSF for controlled downstream capacity. We additionally evaluate a stronger, \emph{task-adapted} (TA) variant of this route, in which the same U-Net++ segmenter is trained from scratch on the HSI (and CS) reconstructions rather than reused off the shelf, giving the most competitive reconstruct-then-segment baseline. These routes instantiate the \emph{lift spectrum}: a fixed classical inverse (shared by the off-the-shelf route and by TA, which thus probes the fixed-physics lift at the strongest possible downstream head), a learned but input-independent static projection (SPIFS), and a content-adaptive attention retrieval (STSF). Parameter budgets: ${\approx}36.8$M for STSF, ${\approx}30.2$M for SPIFS, and ${\approx}29.4$M for the plain image-domain U-Net++; the lift ablation (Table~\ref{tab:lift}) compares retrieval modules within the same STSF network at a matched $36.6$--$36.9$M budget.

\noindent\textbf{Training protocol.} The segmentation experiments here and the reconstruction probe of Sec.~\ref{subsec:benchmark} use independent optimization settings, both specified in full in Supplementary Sec.~S1 (optimizer and schedule, per-dataset epochs and batch sizes, the TPLS phase fractions of Eq.~\eqref{eq:tpls_schedule}, and the SPIFS re-implementation). SPIFS is retrained end-to-end from scratch under the same optimization settings and the same fixed acquisition, without its original two-stage pretraining; fixing the acquisition may understate SPIFS, which was designed to jointly optimize its illumination. Because a shared learning rate can itself favor one arm, we additionally swept the SPIFS learning rate over a $10\times$ range on MNIST, where that baseline is least seed-stable: two settings are strictly worse, and at the baseline's best setting the MNIST margin narrows from $+9.9$ to $+6.5$~pp while remaining several times that setting's cross-seed spread ($1.1$~pp).

\noindent\textbf{Code and models.} Training and evaluation code and pretrained checkpoints are available at \url{https://github.com/Hanyuyuan6/STSF-TPLS}, with the checkpoints hosted on Hugging Face (\url{https://huggingface.co/hanyuyuan/STSF-TPLS-weights}); the reconstruction-probe benchmark of Supplementary Table~S5 is maintained separately at \url{https://github.com/Hanyuyuan6/Comparison-of-SPI-DLs}.

\noindent\textbf{Evaluation metrics.} We report mean intersection-over-union (mIoU) and mean Dice (mDice). Because the foreground occupies a small fraction of each scene, the background IoU saturates near unity and inflates class-averaged scores; we therefore report \emph{foreground} mIoU and mDice (background excluded from the per-class mean) as the primary metrics. Means and margins are computed from unrounded values and rounded once for display.

\subsection{The Lift's First Axis: Selecting a Temporal Encoder}
\label{subsec:benchmark}

The measurements form a 1D vector that can be permuted without information loss when the pattern schedule is known (confirmed empirically by the acquisition-order ablation of Supplementary Sec.~S5), so the design question is not whether the acquisition order carries information but which encoder maps a length-$M$ measurement vector into a spatial feature map most parameter-efficiently.

The lift has two axes; this section charts the first, how the measurement sequence is encoded. We run a controlled, parameter-matched benchmark of inversion pipelines that map SPS measurements to spatial images (per-architecture specifications in Supplementary Sec.~S1), with reconstruction as a task-agnostic probe: fidelity measures how much scene structure each encoder recovers, without tying the comparison to any segmentation head or dataset. The sequence- and graph-based pipelines share one convolutional decoder and vary only in the temporal encoder; the FC, CNN, and U-Net baselines use their native heads at a matched budget. The benchmark comprises a physics-based HSI baseline and eight data-driven architectures: fully connected (FC), convolutional neural network (CNN), U-Net, Transformer, graph convolutional network (GCN), recurrent neural network (RNN), long short-term memory (LSTM), and GRU.

\begin{table}[t]
\centering
\caption{Reconstruction probe used for temporal-encoder selection, at the deployed 3.13\% sampling rate: PSNR (dB, $\uparrow$) and SSIM ($\uparrow$) on MNIST, Fashion-MNIST, and CelebA; parameter count at $M{=}512$. Every cell is a single training run under one shared selection protocol, so the comparison is internally controlled but carries no seed uncertainty (Sec.~\ref{subsec:benchmark}); the full five-condition probe, adding the 6.25\% and 12.5\% MNIST rates, is Supplementary Table~S5. \textbf{Bold}: best per column.}
\label{tab:probe}
\setlength{\tabcolsep}{3.4pt}
\begin{tabular}{l c cc cc cc}
\hline
\multirow{2}{*}{Models} & \multirow{2}{*}{\shortstack{Params\\(M)}} &
\multicolumn{2}{c}{MNIST} & \multicolumn{2}{c}{FMNIST} & \multicolumn{2}{c}{CelebA} \\
\cline{3-8}
 & & PSNR & SSIM & PSNR & SSIM & PSNR & SSIM \\
\hline
Physics-based & N/A & 9.02 & 0.431 & 10.06 & 0.378 & 10.35 & 0.278 \\
FC & 136.4 & 13.88 & 0.699 & 15.94 & 0.611 & 15.08 & 0.437 \\
CNN & 141.1 & 13.37 & 0.697 & 15.40 & 0.606 & 14.88 & 0.434 \\
U-Net & 132.5 & 13.29 & 0.686 & 15.25 & 0.604 & 14.64 & 0.431 \\
GCN & 136.4 & 13.15 & 0.679 & 14.02 & 0.545 & 14.27 & 0.413 \\
Transformer & 146.9 & 14.50 & 0.722 & 15.75 & 0.610 & 14.67 & 0.422 \\
RNN & 135.3 & 13.06 & 0.675 & 15.72 & 0.601 & 13.16 & 0.367 \\
LSTM & 138.5 & 16.05 & 0.765 & 18.25 & 0.684 & 15.67 & \textbf{0.454} \\
GRU & 137.4 & \textbf{16.32} & \textbf{0.768} & \textbf{18.37} & \textbf{0.686} & \textbf{15.69} & 0.453 \\
\hline
\end{tabular}
\end{table}

Table~\ref{tab:probe} shows that the recurrent gated encoders (LSTM/GRU) recover scene structure markedly better than the feed-forward, convolutional, or graph alternatives at a matched budget; in particular the fully connected baseline, which reads the same measurement vector with no sequential inductive bias, is consistently outperformed, so the advantage is architectural rather than a capacity artifact. GRU attains the best value in every reported cell across three sampling rates and three datasets (Supplementary Table~S5), except CelebA SSIM, where LSTM leads by $0.001$. We therefore adopt GRU, the family's most parameter-efficient member. This selection rests on a stated proxy: an encoder that recovers more scene structure also supplies a richer feature map to the downstream lift, so reconstruction fidelity serves as an encoder-selection surrogate, not a guarantee that the reconstruction ranking transfers verbatim to segmentation.

Two protocol notes bound this probe. First, the parameter budgets in Supplementary Table~S5 ($\sim130$--$160$M) belong to this controlled \emph{reconstruction} benchmark alone, not the deployed STSF \emph{segmentation} network ($\approx36.8$M, Sec.~\ref{sec:setup}); the ranking's transfer to the deployed regime is supported by, not proven beyond, the end-to-end results of Sec.~\ref{subsec:sim}. Second, every probe selects its best epoch on the split it reports (split composition in Supplementary Sec.~S1). The same rule applies to all eight architectures, but absolute PSNR/SSIM are optimistic and the single-run ranking has no seed uncertainty; the ranking recurs across all five dataset/rate conditions. The lift's second axis, the retrieval, requires the end-to-end segmentation protocol; the next section fixes it (Table~\ref{tab:lift}).

\subsection{Simulated Comparison and Ablations}
\label{subsec:sim}

\begin{figure}[t]
\centering\includegraphics[width=\linewidth]{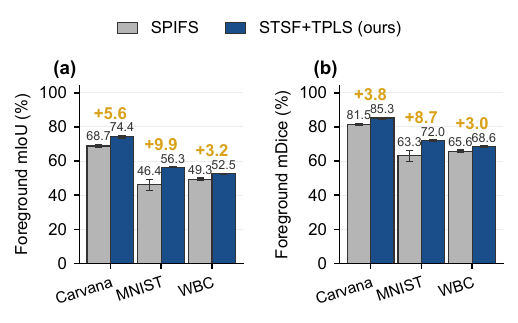}
\caption{Foreground segmentation accuracy of the proposed STSF+TPLS versus the prior single-pixel baseline SPIFS on the three simulated test sets at a 3.13\% sampling rate. (a) Foreground mIoU and (b) foreground mDice, averaged over three training seeds (error bars span the seed min--max); gold labels give the absolute improvement of STSF+TPLS over SPIFS in percentage points (pp). STSF+TPLS leads on every dataset on both metrics, most widely on MNIST.}
\label{fig:main_compare}
\end{figure}

Figure~\ref{fig:main_compare} reports the main comparison. Under the shared, fixed acquisition in simulation, STSF+TPLS outperforms the SPIFS baseline on every dataset, raising the foreground mIoU by $+5.6$, $+9.9$, and $+3.2$ percentage points (pp), to $74.4$, $56.3$, and $52.5$, on Carvana, MNIST, and WBC, and the foreground mDice by $+3.8$, $+8.7$, and $+3.0$~pp (to $85.3$, $72.0$, and $68.6$). The margin is widest on MNIST, whose sparse strokes leave a generic dense lift the least redundancy to fall back on.

\begin{figure}[t]
\centering\includegraphics[width=\linewidth]{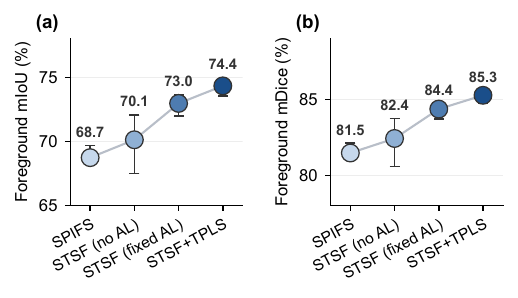}
\caption{Cumulative component ablation on the Carvana test set (foreground metrics): SPIFS $\rightarrow$ STSF (no AL) $\rightarrow$ STSF (fixed AL) $\rightarrow$ STSF+TPLS, adding in turn the STSF architecture, the fixed-weight auxiliary loss, and the TPLS schedule. (a) Foreground mIoU and (b) foreground mDice rise monotonically; markers are three-seed means, whiskers span the seed min--max. The no-AL rung's wide spread versus the tight ones after it shows the auxiliary loss also stabilizes training.}
\label{fig:ablation}
\end{figure}

\begin{table}[t]
\centering
\caption{The 1D-to-2D lift, ablated: each row swaps only the retrieval module inside the same STSF network at a matched budget (36.6--36.9M); foreground mIoU/mDice (\%, mean${}\pm{}$std over three training seeds), WBC test set, 3.13\% sampling. Cross-attention is the most accurate with a tight seed spread. ``-style'' rows are parameter-matched implementations of each family's core primitive inside the STSF lift, not reproductions of the cited systems. \textbf{Bold}: the proposed lift.}
\label{tab:lift}
\setlength{\tabcolsep}{4pt}%
\resizebox{\columnwidth}{!}{%
\footnotesize
\begin{tabular}{l cc c}
\hline
1D$\rightarrow$2D lift & mIoU & mDice & Params \\
\hline
\textbf{Cross-attention (ours)} & \textbf{52.50}$_{\pm0.31}$ & \textbf{68.59}$_{\pm0.27}$ & 36.8M \\
Recurrent broadcast (FSRCNN-style)~\cite{fsrcnn} & 50.86$_{\pm0.26}$ & 67.11$_{\pm0.23}$ & 36.9M \\
Super-res.\ conv.\ (sub-pixel-style)~\cite{espcn} & 50.52$_{\pm0.59}$ & 66.84$_{\pm0.54}$ & 36.6M \\
Implicit field (INR-style)~\cite{nerf} & 49.92$_{\pm1.80}$ & 66.23$_{\pm1.68}$ & 36.8M \\
Kolmogorov--Arnold (KAN-style)~\cite{kan} & 49.46$_{\pm0.50}$ & 65.84$_{\pm0.51}$ & 36.8M \\
State space (Mamba-style)~\cite{mamba} & 48.20$_{\pm1.72}$ & 64.64$_{\pm1.62}$ & 36.9M \\
\hline
\end{tabular}%
}
\end{table}

\looseness=-1 Beyond this headline margin, two ablations attribute the gain to specific design choices. The lift ablation (Table~\ref{tab:lift}) fixes the spectrum's second axis, the 1D-to-2D retrieval: it swaps only the retrieval module inside the same STSF network at a closely matched parameter budget, averaged over three training seeds; cross-attention is the strongest ($52.5{\pm}0.3$~mIoU), ahead of the recurrent-broadcast projection it replaces by $+1.6$~pp and of the implicit-neural-field lift by $+2.6$~pp; the single-seed near-tie with the latter ($+0.35$~pp) reflects its high seed variance (${\pm}1.8$ vs.\ ${\pm}0.3$); the state-space lift ranks lowest on average, by a margin ($1.3$~pp) smaller than its own seed spread (${\pm}1.7$). The component ablation on Carvana (Fig.~\ref{fig:ablation}, three training seeds) instead adds one element at a time and rises monotonically in the three-seed means, with per-step mIoU increments of $+1.4$, $+2.8$, and $+1.4$~pp; only the last step is positive on every seed individually. The auxiliary loss also stabilizes training: the no-AL rung spans $4.6$~pp across seeds against ${\le}1.7$~pp once present, so that step raises accuracy and cuts variance together. Figure~\ref{fig:STSF}(f--i) shows how this supervision shapes the intermediate representation: TPLS, which decays the auxiliary weight, in fact ends marginally \emph{behind} a fixed weight on the auxiliary reconstruction itself ($17.0$ vs.\ $17.7$~dB on the sample shown) while leading on segmentation, so the auxiliary objective is best read as scaffolding for the lift rather than a target in its own right.

The Carvana validation and per-loss curves (Supplementary Fig.~S2) confirm that TPLS changes objective priority without destabilizing optimization. STSF segments a measurement sequence in about 14~ms on an RTX 4090 (batch size 1), timed with the released auxiliary head attached. Validating the 36.8M-parameter network on embedded hardware is future work.

\subsection{Operating Regime: Reconstruction Versus Image-Free Under Stress}
\label{subsec:regime}

\looseness=-1 The headline comparison isolates the image-free family: STSF+TPLS improves on the prior image-free baseline SPIFS (Fig.~\ref{fig:main_compare}). A sharper question is how image-free inference fares against the \emph{strongest} reconstruct-then-segment competitor, the task-adapted (TA) baseline that trains a U-Net++ directly on the classical (HSI/CS) reconstructions. We report this comparison in full: it is not uniformly favorable, and its structure is the contribution.

\begin{figure*}[t]
\centering\includegraphics[width=\linewidth]{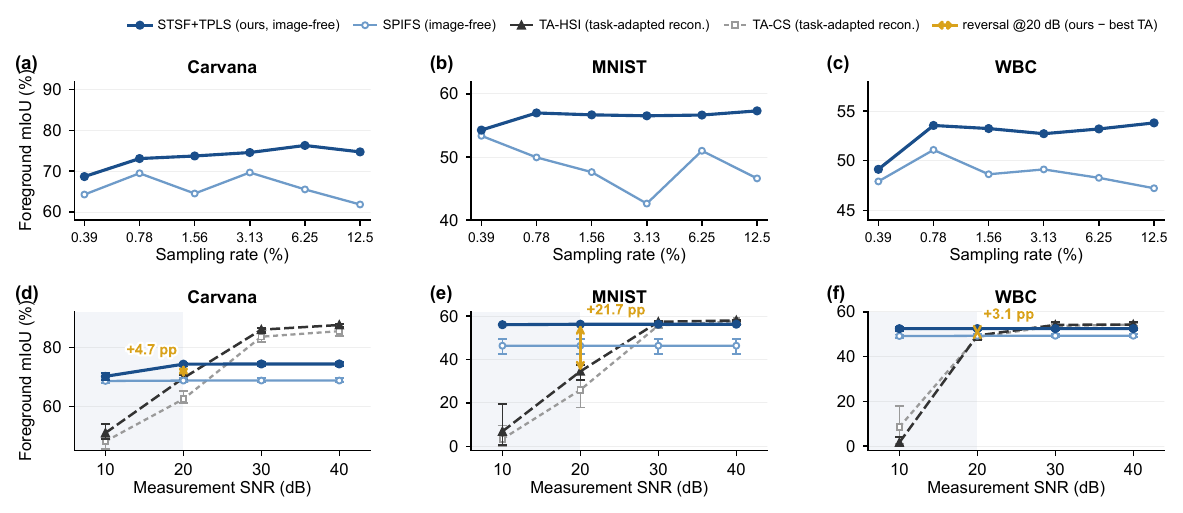}
\caption{Operating regime across sampling rate and measurement noise; foreground mIoU at $128^2$. (a--c) Clean rate sweep (single seed), image-free methods only: STSF+TPLS (deep blue) holds a flat plateau across a $32\times$ span and leads SPIFS (light blue) at every rate, while SPIFS swings non-monotonically; the reconstruction baseline's own rate sweep is tabulated in Supplementary Table~S4. (d--f) Noise sweep at $3.13\%$, calibrated on the AC-coupled bucket; whiskers span the min--max over three training $\times$ three noise seeds per arm. Image-free inference is near-invariant, whereas the task-adapted (TA) baseline collapses below ${\sim}20$~dB (shaded), so image-free overtakes even the strongest clean-trained TA variant at 20~dB by $+4.7$, $+21.7$, and $+3.1$~pp (gold). Reconstruction wins the noiseless limit; image-free wins the noisy, deployment-relevant regime.}
\label{fig:envelope}
\end{figure*}

\noindent\looseness=-1\textbf{Image-free accuracy survives a $32\times$ rate cut.} The rate axis itself is the first argument for the image-free lift (Fig.~\ref{fig:envelope}(a--c); one model trained per rate). STSF+TPLS holds a flat plateau across the full $0.39$--$12.5\%$ sweep, a $32\times$ span: at $0.39\%$ sampling ($64$ measurements, $256\times$ fewer samples than pixels) it still reaches $68.7/54.3/49.1$ foreground mIoU on Carvana/MNIST/WBC, within $5.7/2.0/3.4$~pp of the $3.13\%$ headline of Fig.~\ref{fig:main_compare}, and at $0.78\%$ MNIST and WBC already match it ($57.0$ and $53.5$). It also leads SPIFS at every swept rate on all three datasets. Re-running the narrowest margin (MNIST at $0.39\%$) at three seeds turns the single-seed $+0.9$~pp gap into a $+1.4$~pp mean ($54.3{\pm}0.03$ vs.\ $52.9{\pm}0.5$), worst-case pairing $+0.9$~pp; the remaining rates are single-seed, read as a consistent ordering rather than per-rate effect sizes. SPIFS does not improve monotonically with rate at this seed; the STSF+TPLS masks remain object-shaped down to $0.39\%$ (Supplementary Sec.~S8). The $3.13\%$ rate used throughout is therefore a fixed anchor for the noise study, not a sweet spot the method requires.

\noindent\looseness=-1\textbf{In the clean limit, reconstruction wins.} At a 3.13\% sampling rate without measurement noise, TA is the most accurate method on all three datasets, exceeding STSF+TPLS by $+13.3$ (Carvana), $+1.5$ (MNIST), and $+1.8$~pp foreground mIoU (WBC), means over three training seeds on both sides. Its lead is largest on Carvana, whose smooth silhouettes the classical reconstruction renders well enough for the \emph{retrained} segmenter to exploit, and it persists down to $0.39\%$ sampling (Carvana TA-HSI $87.9\!\rightarrow\!81.9$; Supplementary Table~S4). A noiseless benchmark would therefore call reconstruction, not image-free inference, the method of choice.

\begin{figure*}[t!]
\centering\includegraphics[width=\linewidth]{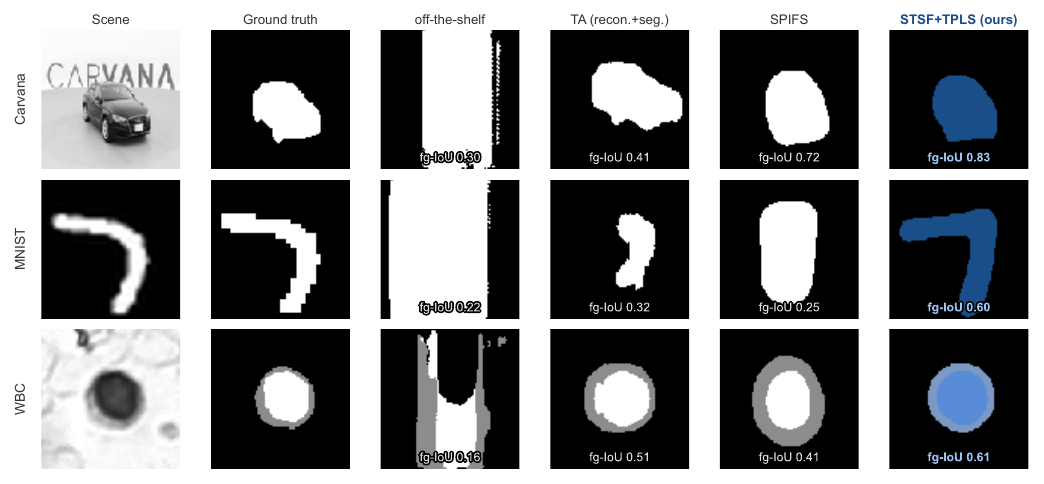}
\caption{The reversal, qualitatively: segmentations at a 20~dB measurement SNR ($3.13\%$ sampling; noise seed~1 and training seed~42 under the AC calibration of Sec.~\ref{subsec:regime}, each pipeline drawing its own realization; per-sample foreground IoU on each mask). The \emph{off-the-shelf} U-Net++, trained on fully-sampled images, collapses on the speckle-corrupted reconstruction (test-set foreground mIoU $29.8$/$27.6$/$21.3$ on Carvana/MNIST/WBC); \emph{task-adapted} (TA) retraining of the same head degrades to $70.1$/$35.9$/$47.9$; SPIFS settles into its usual rounded blob ($69.7$/$42.6$/$49.1$); STSF+TPLS stays highest ($74.5$/$56.5$/$52.8$). The reconstruction the two reconstruct-then-segment heads read is banded even when clean and buried in speckle at this SNR (Supplementary Sec.~S3); Supplementary Fig.~S4 shows the noiseless counterpart, the regime where TA leads on test-set means.}
\label{fig:qual20}
\end{figure*}

\noindent\looseness=-1\textbf{Measurement noise reverses the ranking.} That conclusion is an artifact of noiseless simulation. We evaluate robustness to additive measurement-domain noise calibrated on the AC-coupled bucket signal a real detector produces, $\sigma=\mathrm{std}(s_{\mathrm{AC}})\cdot 10^{-\mathrm{SNR}/20}$, at SNR${}\in\{40,30,20,10\}$~dB (Fig.~\ref{fig:envelope}). Both families read measurements corrupted at the \emph{same} SNR under this calibration, yet respond oppositely. Image-free inference is near-invariant across the noise axis (STSF+TPLS foreground mIoU on Carvana $74.4\!\rightarrow\!74.4\!\rightarrow\!74.3\!\rightarrow\!70.2$ over $40/30/20/10$~dB, three-seed mean; MNIST and WBC flat to within $0.3$~pp), whereas the task-adapted accuracy falls steeply below ${\sim}20$~dB (Carvana TA-HSI $87.5\!\rightarrow\!69.6\!\rightarrow\!51.0$ at $40/20/10$~dB, three-seed means; the CS variant adds an ADMM-$\ell_1$ prior but inherits the same shift). At 20~dB the ranking reverses: image-free inference overtakes even the strongest clean-trained TA variant by $+4.7$ (Carvana), $+21.7$ (MNIST), and $+3.1$~pp (WBC). With three training seeds on every arm, the collapse is seed-robust (MNIST TA $34.6{\pm}2.8$) and the worst-case seed pairing does not reverse on any dataset ($+2.9$, $+18.6$, and $+0.2$~pp). Figure~\ref{fig:qual20} shows what this reversal looks like, down to the speckle-buried reconstruction the TA segmenter must read (Supplementary Fig.~S3).

\noindent\textbf{Why noise reverses the ranking.} The collapse is not a conditioning effect. The shared Hadamard adjoint is a well-conditioned scaled isometry ($\Phi\Phi^{\!\top}\!=\!N\mathbf{I}$, with $N{=}HW$ the pixel count), so per measured mode the reconstruction is no noisier than the measurements. What differs is where each pipeline normalizes, and that placement sets the relative perturbation $\delta$ its segmenter reads, the per-coordinate RMS perturbation of the input divided by that input's min--max range. To first order, the ratio factorizes as
\begin{equation}
\frac{\delta_{\mathrm{TA}}}{\delta_{\mathrm{IF}}}
=\underbrace{\frac{\sqrt{M}}{N}}_{\text{dimension}}\times\underbrace{\frac{R_{\mathrm{IF}}}{R_{\mathrm{TA}}}}_{\text{range}}.
\label{eq:amplify_main}
\end{equation}
\looseness=-1 The dimension factor $\sqrt{M}/N\!\approx\!1.4\times10^{-3}$ favors the reconstruction, whose well-conditioned adjoint dilutes the noise energy over $N$ pixels. The range factor overwhelms it. Image-free decoding normalizes the raw measurement vector, whose range is anchored by the large DC coefficient, so the AC-referenced noise stays a small \emph{absolute} perturbation of the network input. The reconstruction receives the same DC, but as a spatially uniform plane, invisible to the shift-invariant per-image min--max, so its range is set by the AC image content together with the full-frame noise speckle; measured directly, $R_{\mathrm{IF}}$ exceeds $R_{\mathrm{TA}}$ by roughly four orders of magnitude on these scenes (${\approx}8.9{\times}10^{3}$ on MNIST and ${\approx}4.2{\times}10^{4}$ on Carvana; the DC-to-AC ratio alone contributes ${\approx}21$ and ${\approx}228$). Measured end to end, the same injected measurement noise produces a $20$--$70\times$ larger normalized relative perturbation at the TA segmenter input at 20~dB ($3.5$--$6.7\%$ across the three datasets versus ${\leq}0.4\%$ image-free; derivation and measurement in Supplementary Sec.~S3). The clean-trained head is thereby driven out of distribution, and overlap geometry punishes this hardest where the foreground is sparse, consistent with MNIST falling steepest. This account is falsifiable, and we tested it: retraining the TA head on noise-augmented reconstructions restores accuracy at the trained SNR and degrades away from it (Table~\ref{tab:naug}; protocol in Supplementary Sec.~S3), so the collapse is clean-training distribution shift, not an information limit. Direct image-free decoding never forms that image and remains much closer to its clean training distribution across the tested noise grid without noise-specific retraining. We claim this reversal only within the tested noise range and calibration.

\begin{table}[t]
\centering
\caption{Noise-augmented task-adapted retraining on MNIST (foreground mIoU, mean over three noise seeds). Retraining the TA head on noisy reconstructions restores accuracy at the trained SNR (\textbf{bold}), confirming the collapse is clean-training distribution shift rather than an information limit, but requires noise-aware training and degrades away from that support; image-free STSF+TPLS stays flat across the tested grid without noise-specific training. Spread is max$-$min across the grid. STSF+TPLS: three-training-seed mean; TA rows: one training seed each; protocol in Supplementary Sec.~S3.}
\label{tab:naug}
\setlength{\tabcolsep}{4.5pt}%
\begin{tabular}{lcccccc}
\hline
Eval SNR (dB) & 40 & 30 & 20 & 15 & 10 & Spread \\
\hline
Clean-trained TA & 57.9 & 57.8 & 36.0 & 13.2 & 0.6 & 57.3 \\
TA, noise-aug @\,20~dB & 55.7 & 57.9 & \textbf{57.8} & 49.1 & 33.5 & 24.4 \\
TA, noise-aug mixed & 57.3 & 57.9 & 57.3 & 55.4 & 49.9 & 8.0 \\
STSF+TPLS (image-free) & 56.3 & 56.3 & 56.3 & 56.2 & 56.1 & 0.2 \\
\hline
\end{tabular}
\end{table}

\noindent\looseness=-1\textbf{The spectrum orders the transition.} These behaviors follow the lift spectrum: the fixed-physics TA is sharpest when clean but meets an out-of-distribution reconstruction under noise, the content-adaptive STSF trades a little clean accuracy for a stable lift, and the learned-static SPIFS sits between them, flat but capped. The map shows \emph{where} each design wins, not that image-free is universally better, anticipating the finite-SNR hardware regime of Sec.~\ref{subsec:phy}. It also supports a minimax-regret reading: across the tested noise grid the content-adaptive lift never trails the leader by more than the clean-limit gap ($13.3$~pp, Carvana), whereas the clean-trained reconstruction route falls farther behind at the noisy end of every dataset ($19.2$~pp on Carvana at $10$~dB, farther still on MNIST and WBC; Fig.~\ref{fig:envelope}(d--f))---so when SNR is unknown but lies within the tested grid, the content-adaptive lift minimizes worst-case regret in every dataset tested.

\FloatBarrier
\subsection{Failure Modes Across the Lift Spectrum}
\label{subsec:failure}

\begin{figure}[t]
\centering\includegraphics[width=\linewidth]{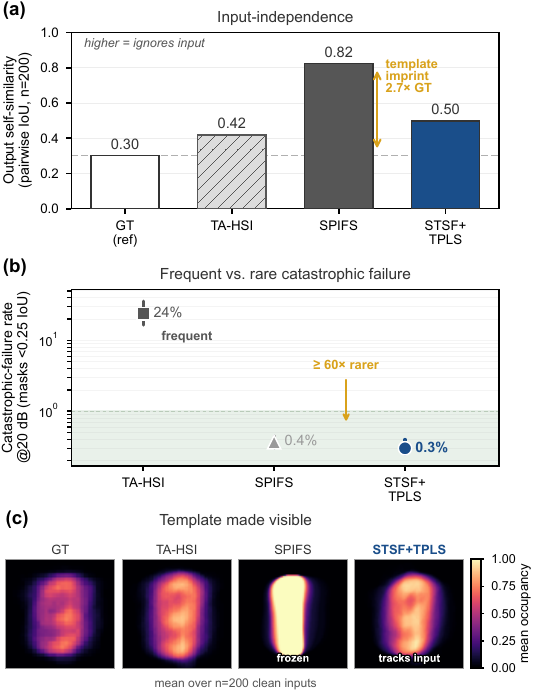}
\caption{Quantitative failure signatures across the lift spectrum (MNIST, $128^2$, 3.13\% sampling; per-example gallery in Supplementary Fig.~S6). (a) Output self-similarity, the pairwise IoU among predictions over $n{=}200$ inputs (higher ${=}$ more input-independent): SPIFS ($0.82$) far exceeds the intrinsic scene self-similarity ($0.30$, dashed), evidencing a template (STSF+TPLS $0.50$, TA-HSI $0.42$). (b) Catastrophic-failure rate at 20~dB, the fraction of the $2000$ test masks below $0.25$ foreground IoU (log scale, three training seeds): TA-HSI fails on $24\%$ of scenes versus $0.4\%$ (SPIFS) and $0.3\%$ (STSF+TPLS). (c) Mean predicted occupancy over the same $n{=}200$ clean inputs as (a): SPIFS's mean is the template itself; TA-HSI and STSF+TPLS track the diffuse ground-truth occupancy.}
\label{fig:failure}
\end{figure}

The regime map leaves a second question. When each design fails, \emph{how} does it fail? On the MNIST stress test, inspecting predictions as the acquisition degrades reveals three characteristic failure signatures, one per region of the lift spectrum (Fig.~\ref{fig:failure}). These are dominant tendencies rather than exclusive modes, quantified per mode below.

\noindent\looseness=-1\textbf{Collapse (fixed-physics lift).} The reconstruction path (TA) fails not by confidently drawing the wrong thing but by ceasing to draw. Driven out of distribution by the enlarged normalized task-input perturbation of Sec.~\ref{subsec:regime}, its segmentation head starves: of the $24.0\pm10.4\%$ of scenes it fails at 20~dB (foreground IoU below $0.25$; three training seeds), $94\%$ retain under $0.70$ of the ground-truth foreground, the median failing mask holding only $0.38\pm0.11$ of it. The stroke thins to a placed but ink-starved residue or shatters into fragments (Supplementary Fig.~S6), and on $6.0\pm8.5\%$ of scenes it blanks out altogether (below $5\%$ predicted foreground). Only $5.9\pm5.5\%$ of its failures are confident, full-ink wrong shapes; the dominant observed failure is therefore foreground starvation rather than false-structure retention. On hardware this reads as instability (Sec.~\ref{subsec:phy}).

\noindent\looseness=-1\textbf{Imprinting (learned-static lift).} SPIFS degenerates to a near input-independent template. Across 200 different test scenes its predictions resemble one another (pairwise IoU $0.82$) far more than the scenes themselves do ($0.30$; Fig.~\ref{fig:failure}(a)), and its accuracy barely moves under measurement noise (MNIST foreground mIoU $42.63$ at training seed 42, shifting under $0.02$~pp from 40 to 10~dB): the static projection has learned an average mask rather than a mapping from the input; its mean prediction is a single saturated template (Fig.~\ref{fig:failure}(c)).

\noindent\looseness=-1\textbf{Coarsening (content-adaptive lift).} STSF+TPLS produces coarse but correctly placed masks that preserve topology (the loop of the ``0'' and both loops of the ``8'' survive). Its output tracks the input more than SPIFS (self-similarity $0.50$, between the scene floor and the SPIFS template) yet stays stable: accuracy varies by only $0.47$~pp across training seeds versus $6.68$~pp for SPIFS, a ${\approx}14\times$ tighter spread. Faced with uncertainty it dilates into a conservative envelope, the same shape-faithful over-segmentation seen on the hardware car target (Fig.~\ref{fig:exp}(c)), rather than collapsing or imprinting. Across the $2000$ MNIST test scenes at 20~dB only $0.3\%$ of its masks fall below $0.25$ foreground IoU, versus $24\%$ for the reconstruction path (three-seed means; Fig.~\ref{fig:failure}(b)); even these rare failures keep drawing, over-full rather than absent, carrying a median $2.3\times$ the ground-truth foreground.

In sum, the fixed-physics route collapses, the learned-static route imprints, and the content-adaptive route coarsens---three observed signatures that explain the regime map of Sec.~\ref{subsec:regime}.

\subsection{Physical Experiments}
\label{subsec:phy}

\begin{figure*}[t]
\centering
\includegraphics[width=\linewidth]{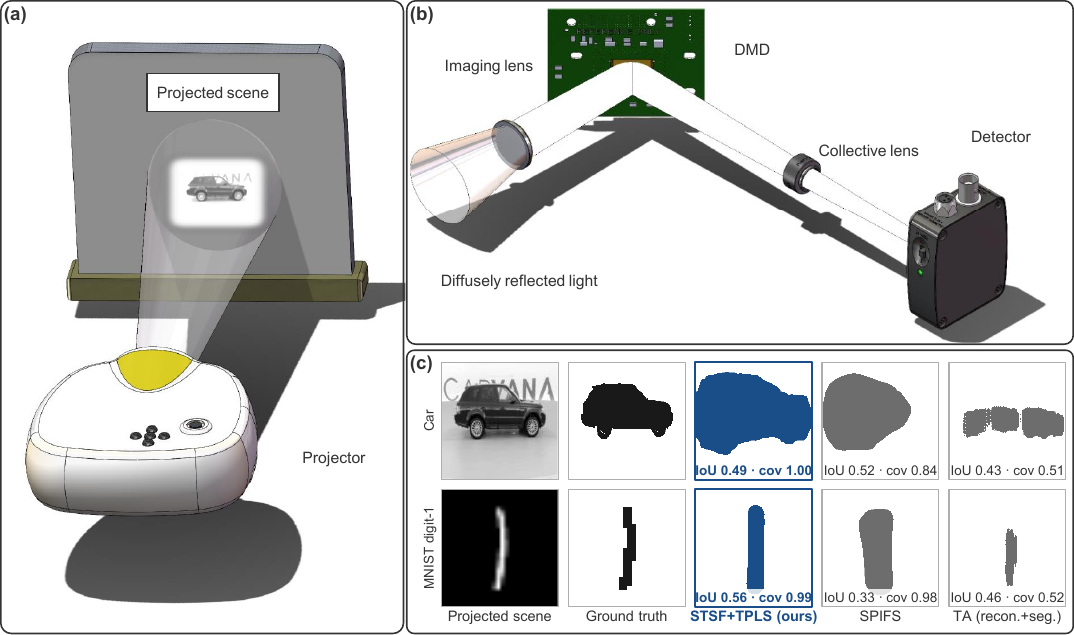}
\caption{Real single-pixel bench, two valid physical targets at 3.13\% sampling. (a,~b) Acquisition: a projector displays a held-out test image on a matte screen; the diffusely reflected light is imaged onto a DMD applying $\pm1$ natural-order Hadamard patterns and focused onto a bucket detector, which records 512 coded measurements. (c) Predicted masks for both targets, each carrying Mask IoU and foreground coverage (cov). \emph{Car}, a projected scene without registered ground truth, read qualitatively: STSF+TPLS encloses the whole vehicle (cov $1.00$) but is dilated, which Mask IoU penalizes (IoU $0.49$), while SPIFS settles into a rounded blob and TA fragments the silhouette (cov $0.51$). \emph{MNIST digit-1}, displayed at a fixed projector position from a known test-set image, so its ground-truth mask is known a priori: STSF+TPLS recovers the bar (IoU $0.56$, cov $0.99$); SPIFS (IoU $0.33$) and the task-adapted baseline (TA, IoU $0.46$, cov $0.52$) do not. The digit values are for the first of the two acquisitions of this target; Fig.~\ref{fig:phyrev} averages both.}
\label{fig:exp}
\end{figure*}

\begin{figure}[t]
\centering
\includegraphics[width=\linewidth]{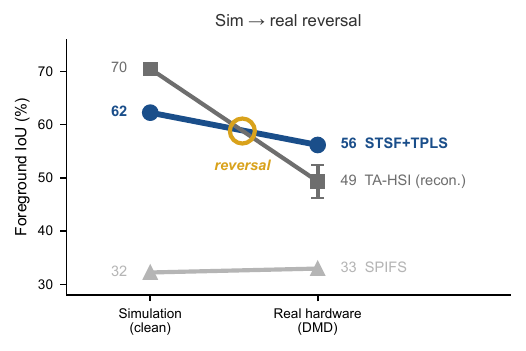}
\caption{Sim-to-real on the physical bench: MNIST digit-1, the reliable-ground-truth target of Fig.~\ref{fig:exp}(c), at 3.13\% sampling. The task-adapted route (TA), most accurate in clean simulation ($0.70$), collapses below the image-free STSF+TPLS on hardware ($0.49$, two-run mean) and swings between repeated acquisitions ($0.47$ run-to-run IoU), whereas STSF+TPLS stays highest and nearly reproduces its own mask ($0.99$). This is the hardware signature of the noise reversal of Sec.~\ref{subsec:regime}, a proof of concept on one target that corroborates rather than establishes the simulated result.}
\label{fig:phyrev}
\end{figure}

\looseness=-1 The experimental configuration is shown in Fig.~\ref{fig:exp}(a,b). A digital projector (JMGO N1 Air, triple-color laser, 1080p) illuminates a matte projection screen to generate scenes drawn from the held-out test set. The diffusely reflected light is collected by relay optics and directed onto a DMD (Texas Instruments DLP7000, $1024{\times}768$ micromirrors at a $13.68~\mu$m pitch, on a ViALUX SuperSpeed V-Module supporting full-frame binary pattern rates up to $22{,}727$~Hz). The DMD temporally encodes the incoming optical field using a precomputed sequence of Hadamard patterns; for each pattern, a single-pixel detector (Thorlabs PDA36A2, silicon, 350--1100~nm) records the total reflected intensity.
Because the DMD is binary, each $\pm1$ Hadamard row is displayed as a complementary pattern pair and the two detector readings are subtracted; the recorded measurement sequence is therefore already differenced and directly realizes the $\pm1$ acquisition model of Eq.~\eqref{eq:sps_forward}. All measurements are acquired under identical illumination and optical settings to ensure consistency across methods.

Figure~\ref{fig:exp}(c) shows representative results at 3.13\% sampling. The bench is a proof of concept; the quantitative claims rest on simulation.\looseness=-1{} The trained network runs end-to-end on real measurements without fine-tuning (confirming sim-to-real transfer of the image-free lift) and exposes an evaluation problem a simulation-only study would miss: under real acquisition every method degrades toward a coarse mask, and on the target without a registered ground truth, region-overlap metrics stop ranking methods meaningfully; Mask IoU conflates under- and over-segmentation and is biased by object size \cite{boundary_iou, metrics_reloaded}, penalizing the shape-faithful but visibly dilated STSF+TPLS car mask (cov $1.00$). On the one target with a reliable ground truth, the MNIST digit-1 bar, projected at a fixed position from a known test-set image so its mask is known a priori, image-free leads quantitatively (Mask IoU $0.56$ vs.\ $0.49$ for the task-adapted baseline and $0.33$ for SPIFS, two-run means), and across those two acquisitions its mask is nearly reproduced (run-to-run IoU $0.99$) while the TA mask swings ($0.47$), the hardware signature of the noise sensitivity of Sec.~\ref{subsec:regime} (Fig.~\ref{fig:phyrev}). We read the bench through run-to-run stability and the reliable-ground-truth target rather than overlap ranking; a registered-ground-truth benchmark is future work.

\section{Discussion and Conclusion}

This work identifies the 1D-to-2D lift as the central design axis of image-free SPS. STSF uses learned queries to retrieve content-dependent evidence directly from the measurement sequence into task features. TPLS schedules a reconstruction loss through a parallel auxiliary head: that loss shapes the shared representation during training, while the reconstructed output is not consumed by the segmentation head and the branch can be pruned for deployment. Mask prediction therefore uses content-adaptive direct lifting, not learned image reconstruction followed by segmentation.

\looseness=-1 In simulation at a 3.13\% sampling rate, STSF+TPLS improves foreground mIoU over SPIFS by $+3.2$ to $+9.9$ percentage points on three datasets and remains competitive down to $0.39\%$ sampling; it also transfers without fine-tuning to real single-pixel hardware as a proof of concept. The deeper result is the clean-to-noisy reversal. The task-adapted fixed-physics reconstruction route is most accurate without measurement noise, but at 20~dB its normalized relative task-input perturbation is 20--70$\times$ larger than that of direct image-free inference, and the ranking reverses, in simulation and on the physical bench. Under acquisition stress, the fixed-physics route collapses, the learned-static route imprints, and the content-adaptive route coarsens. Thus, within the tested fixed acquisition, datasets, and noise ranges, measurement-to-space adaptivity organizes both the operating envelope and its failure signatures. STSF and TPLS instantiate the resulting design principle: use content-adaptive lifting for inference, and use learned reconstruction only as scheduled training supervision. Extending this principle beyond segmentation and conditioning the lift on changing sensing operators are the next tests.

\noindent\looseness=-1\textbf{Limitations and future work.} The headline comparison, the noise sweep, the lift ablation, and the component ablation carry three-seed statistics; the remaining studies (the reconstruction probe, the rate sweep, the acquisition-order and shot-noise sweeps, the noise-augmented TA retraining, and the gradient measurement, the last averaged over five repeats) are single-seed and disclosed as such where reported. The evaluated scenes contain a single dominant object, reflecting the 512-measurement information budget. $\Phi$ is fixed and is not a network input, so any change in sampling rate, pattern set, or resolution requires retraining, unlike unrolled methods that embed $\Phi$; conditioning the lift on $\Phi$ is a natural extension. The Gaussian noise study is confined to the tested SNR grid and calibration, and the reversal further survives a physically faithful shot-noise (Poisson) sweep down to $10^{2}$ photons per measurement (Supplementary Sec.~S4); detector nonlinearity is left to future work.

\section*{Acknowledgment}
This work was supported in part by the Key Research and Development Program of Shaanxi Province under Grant 2024CY2-GJHX-89, in part by the Fundamental Research Funds for the Central Universities, and in part by the National Natural Science Foundation of China under Grant 62375215.

\bibliographystyle{IEEEtran}
\bibliography{cas-refs}

\iflatexml
  \setcounter{section}{0}\setcounter{figure}{0}
  \setcounter{table}{0}\setcounter{equation}{0}
  \renewcommand{\thesection}{S\arabic{section}}
  \renewcommand{\thesubsection}{S\arabic{section}.\arabic{subsection}}
  \renewcommand{\thefigure}{S\arabic{figure}}
  \renewcommand{\thetable}{S\arabic{table}}
  \renewcommand{\theequation}{S\arabic{equation}}
  \makeatletter
  \renewcommand{\thesection@ID}{SI\@section@ID}
  \makeatother
  \section*{Supplementary Material}

\section{Architectural specifications and training protocols}
\label{sec:archspec}

\subsection{Notation and experimental protocol}

Let \(\mathbf{x}\in\mathbb{R}^{N}\) denote the vectorized target scene, and let \(\mathbf{s}\in\mathbb{R}^{M}\) denote the corresponding one-dimensional (1D) measurement vector (i.e., the single-pixel measurement sequence). In our experiments, \(N=128\times128=16{,}384\) and \(M=512\). The sampling ratio is defined as
\begin{equation}
\rho \triangleq \frac{M}{N} = \frac{512}{128^2} = 0.03125 \approx 3.13\%.
\end{equation}
Unless otherwise stated, all models take \(\mathbf{s}\) as input and output a single-channel reconstruction \(\hat{\mathbf{x}}\), with pixel intensities constrained to \([0,1]\) via a sigmoid activation at the network output.

The reconstruction-probe experiments described in this section (Supplementary Table~\ref{tab:encoder_probe}) were run on a workstation equipped with an NVIDIA GeForce RTX 4090 GPU, 128~GB of RAM, and an Intel Core i9-13900K CPU. The image-free segmentation experiments of the main text were instead trained on rented cloud instances, each equipped with an NVIDIA GeForce RTX~4080~SUPER (32~GB), under PyTorch 2.7.0 and CUDA 12.8. All dataset images were resized to \(128\times128\) pixels prior to training. The optimization settings below apply only to the reconstruction probe (the temporal-encoder benchmark, Supplementary Table~\ref{tab:encoder_probe}) and are independent of the segmentation protocol described below (Sec.~\ref{sec:segprotocol}). We trained all models with the AdamW optimizer (weight decay of 0.01) under automatic mixed precision, using a composite reconstruction loss (L1, MSE, and SSIM terms weighted 0.5, 0.3, and 0.2, the SSIM term implemented in its DSSIM form \((1-\mathrm{SSIM})/2\)) and a cosine-annealing learning-rate schedule with an initial learning rate of \(1\times10^{-4}\) and period \(T_{\max}=50\) epochs. Training was performed for 25 epochs with a batch size of 32; only the first half of the cosine period is therefore traversed, and the final learning rate is approximately half the initial value. Each probe selects its best epoch on the split it reports: the torchvision test split for MNIST and Fashion-MNIST, and the held-out 10\% split for CelebA. Each split is disjoint from training, but no dataset holds out a third, never-touched test set, so the absolute PSNR/SSIM of Supplementary Table~\ref{tab:encoder_probe} are optimistic throughout. For consistency, tensor shapes are reported using the \((B,C,H,W)\) convention, where \(B\) denotes the batch size.

We evaluated eight representative learning-based encoders: fully connected (FC), convolutional neural network (CNN), U-Net, Transformer, graph convolutional network (GCN), and three recurrent variants, namely recurrent neural network (RNN), long short-term memory (LSTM), and gated recurrent unit (GRU). Throughout, batch normalization (BN), layer normalization (LN), rectified linear unit (ReLU), convolution (Conv), transposed convolution (ConvT), and multi-layer perceptron (MLP) are used as standard building blocks. Figure~\ref{fig:overview} gives a single overview of the acquisition stage and every evaluated pipeline: panel (a) shows the single-pixel acquisition producing the 1D measurement sequence, panel (c) the classical Hadamard (HSI) reconstruction used by the reconstruct-then-segment baselines, and panels (b, d--k) the eight learning-based encoders, with the shared convolutional decoder defined once in panel (f). Table~\ref{tab:io} complements the diagram with the per-block input/output tensor specifications of each encoder; the subsections below give the corresponding prose specifications.

\begin{figure*}[ht]
\centering
\includegraphics[width=\linewidth]{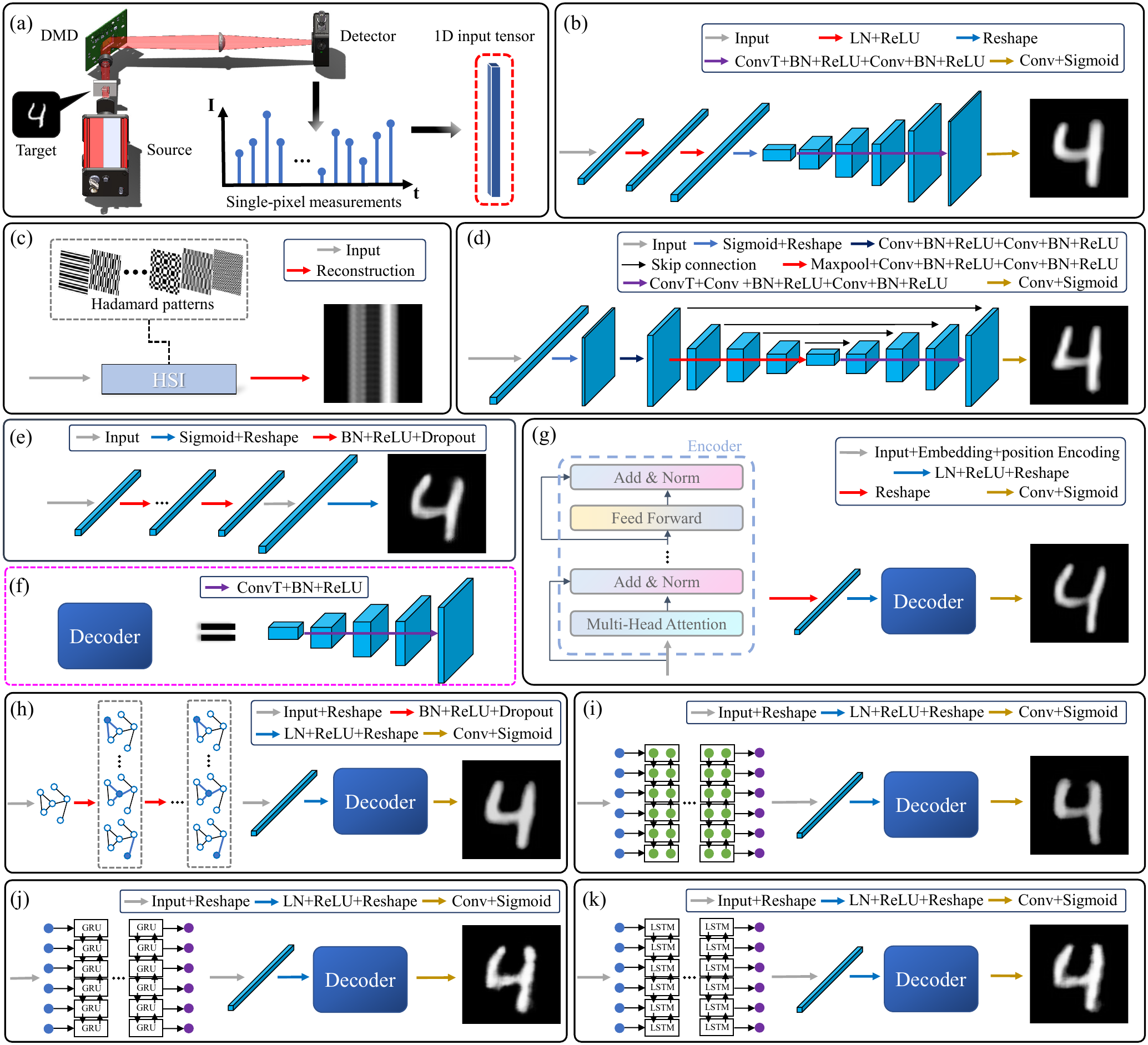}
\caption{Overview of the acquisition stage and all evaluated reconstruction pipelines of the temporal-encoder probe. (a)~Single-pixel acquisition: a DMD modulates the scene with Hadamard patterns and a bucket detector records the 1D measurement sequence, which is the sole input to every method. (b)~CNN-based model. (c)~Classical Hadamard single-pixel imaging (HSI) reconstruction, the physics-based baseline shared by the reconstruct-then-segment pipelines. (d)~U-Net-based model. (e)~FC baseline. (f)~The shared convolutional decoder (learned upsampling head) reused by the Transformer, GCN, and recurrent encoders. (g)~Transformer-based model. (h)~GCN-based model. (i)~RNN-, (j)~GRU-, and (k)~LSTM-based models, which differ only in the recurrent cell. Arrow legends within each panel specify the layer compositions.}
\label{fig:overview}
\end{figure*}

\begin{table*}[ht]
\centering
\caption{Per-block input/output tensor specifications of the eight probe encoders at the paper setting ($M{=}512$ measurements, $128^{2}$ scenes; batch dimension $B$; parameter counts at $M{=}512$). The Transformer, GCN, and recurrent encoders share one convolutional decoder (bottom block; Fig.~\ref{fig:overview}(f)), and their measurement-to-feature heads share the same output interface $(B,32,16,16)$.}
\label{tab:io}
\resizebox{\textwidth}{!}{%
\begin{tabular}{l l l}
\hline
Architecture (Params) & Block & Tensor flow \\
\hline
FC (136.4\,M) & $5\times$[Linear${+}$BN${+}$ReLU] at width $4096$ (dropout $0.2$ after the first four) & $(B,512)\rightarrow(B,4096)$ \\
 & output head: Linear${+}$Sigmoid, then reshape & $(B,4096)\rightarrow(B,16384)\rightarrow(B,1,128,128)$ \\
\hline
CNN (141.1\,M) & measurement MLP: Linear${+}$LN${+}$ReLU ($512{\to}7500{\to}16384$), reshape to seed & $(B,512)\rightarrow(B,1024,4,4)$ \\
 & $5\times$[ConvT$_{4\times4,s2}$${+}$BN${+}$ReLU${+}$Conv$_{3\times3}$${+}$BN${+}$ReLU], channels $1024{\to}32$ & $(B,1024,4,4)\rightarrow(B,32,128,128)$ \\
 & head: Conv$_{3\times3}$${+}$Conv$_{1\times1}$${+}$Sigmoid & $(B,32,128,128)\rightarrow(B,1,128,128)$ \\
\hline
U-Net (132.5\,M) & lift: Linear${+}$Sigmoid, reshape to image grid & $(B,512)\rightarrow(B,1,128,128)$ \\
 & encoder: DoubleConv${+}$ $4\times$[MaxPool${+}$DoubleConv], channels $128{\to}2048$ & $(B,1,128,128)\rightarrow(B,2048,8,8)$ \\
 & decoder: $4\times$[ConvT$_{2\times2,s2}$${+}$skip-concat${+}$DoubleConv] & $(B,2048,8,8)\rightarrow(B,128,128,128)$ \\
 & head: Conv$_{1\times1}$${+}$Sigmoid & $(B,128,128,128)\rightarrow(B,1,128,128)$ \\
\hline
Transformer (146.9\,M) & tokenize ($32$ tokens of length $16$), Linear embed ${\times}\sqrt{512}$ ${+}$ positional enc. & $(B,512)\rightarrow(B,32,16)\rightarrow(B,32,512)$ \\
 & $6\times$TransformerEncoderLayer ($d{=}512$, $16$ heads, FFN $1024$, dropout $0.1$) & $(B,32,512)\rightarrow(B,32,512)$ \\
 & flatten ${+}$ MLP head: Linear($16384{\to}8192$)${+}$LN${+}$ReLU, reshape & $(B,32,512)\rightarrow(B,32,16,16)$ \\
\hline
GCN (136.4\,M) & node partition ($32$ nodes of $16$ features), Linear projection & $(B,512)\rightarrow(B,32,16)\rightarrow(B,32,512)$ \\
 & $8\times$[GraphConv${+}$BN${+}$ReLU] on a learned row-normalized adjacency $(32{\times}32)$ & $(B,32,512)\rightarrow(B,32,512)$ \\
 & flatten ${+}$ MLP head (as above), reshape & $(B,32,512)\rightarrow(B,32,16,16)$ \\
\hline
RNN / LSTM / GRU & sequence partition ($32$ steps of length $16$), Linear projection & $(B,512)\rightarrow(B,32,16)\rightarrow(B,32,256)$ \\
(135.3 / 138.5 / 137.4\,M) & $3$-layer bidirectional cell (hidden $256$ per direction, dropout $0.2$) & $(B,32,256)\rightarrow(B,32,512)$ \\
 & flatten ${+}$ MLP head (as above), reshape & $(B,32,512)\rightarrow(B,32,16,16)$ \\
\hline
Shared decoder & $3\times$[ConvT$_{4\times4,s2}$${+}$BN${+}$ReLU], channels $32{\to}16{\to}8{\to}4$ & $(B,32,16,16)\rightarrow(B,4,128,128)$ \\
(Fig.~\ref{fig:overview}(f)) & head: Conv$_{3\times3}$${+}$Conv$_{1\times1}$${+}$Sigmoid & $(B,4,128,128)\rightarrow(B,1,128,128)$ \\
\hline
\end{tabular}%
}
\end{table*}

\subsection{Fully connected baseline}

The FC baseline (Fig.~\ref{fig:overview}(e)) takes \(\mathbf{s}\) as input and comprises five hidden fully connected layers, each with 4,096 neurons, followed by an output layer. Each hidden layer is followed by BN and ReLU, with dropout (rate of 0.2) additionally applied after the first four. The output layer maps the latent representation to a 16,384-dimensional vector, which is reshaped into a \(128\times128\) image; a sigmoid activation yields normalized pixel intensities.

\subsection{CNN-based model}

The CNN architecture (Fig.~\ref{fig:overview}(b)) reconstructs a normalized \(128\times128\) image from the 1D measurement vector via (i) an input projection module and (ii) a ConvT-based decoder. The input projection comprises two fully connected layers with interleaved LN and ReLU activations, expanding \(\mathbf{s}\) to a 16,384-dimensional latent vector. This vector is reshaped to \((B,1024,4,4)\) and used as the decoder input. The decoder performs progressive learned upsampling using ConvT layers (kernel size of 4, stride of 2, padding of 1), interleaved with \(3\times3\) Conv refinement blocks. Each upsampling stage doubles the spatial resolution while typically reducing the channel dimension. After five upsampling stages, a \(3\times3\) Conv reduces the feature maps to a single channel, followed by a \(1\times1\) Conv and a final sigmoid activation to produce the reconstructed image.

\subsection{U-Net-based model}

The U-Net architecture (Fig.~\ref{fig:overview}(d)) maps \(\mathbf{s}\) to a normalized \(128\times128\) image through an initial fully connected projection followed by a symmetric encoder--decoder with skip connections. The input projection is implemented using a single fully connected layer that expands \(\mathbf{s}\) to 16,384 dimensions; a sigmoid activation bounds the projected values in \([0,1]\). The projected vector is reshaped to \((B,1,128,128)\) and provided as the U-Net input. The encoder contains four downsampling stages to increase representational capacity while reducing spatial resolution; the decoder symmetrically applies four learned upsampling stages and fuses features with corresponding encoder activations via skip connections. A final \(1\times1\) Conv projects the decoder features to a single grayscale channel, and a sigmoid activation produces the reconstructed image.

\subsection{Shared convolutional decoder (learned upsampling head)}

The shared decoder (Fig.~\ref{fig:overview}(f)) implements a three-stage learned upsampling pipeline that maps an input tensor of shape \((B,32,16,16)\) to a normalized \(128\times128\) image. Each stage applies a ConvT layer (kernel size of 4, stride of 2, padding of 1) followed by BN and ReLU, thereby doubling the spatial resolution while reducing the channel dimension (\(32\rightarrow16\rightarrow8\rightarrow4\)). After the final ConvT layer, features are projected to a single channel using a \(3\times3\) Conv followed by a \(1\times1\) Conv; a concluding sigmoid activation constrains the reconstructed intensities to \([0,1]\).

\subsection{Transformer-based model}

The Transformer-based architecture (Fig.~\ref{fig:overview}(g)) maps \(\mathbf{s}\) to a normalized \(128\times128\) image through three stages: patch embedding, Transformer encoding, and convolutional decoding. Let \(M\) denote the measurement length (\(M=512\) in our setting). We partition \(\mathbf{s}\) into \(P\) contiguous, non-overlapping patches, where \(P\) is selected as the largest divisor of \(M\) not exceeding 32 to ensure a uniform patch length \(L=M/P\) (thus \(P=32\) and \(L=16\) for \(M=512\)). Each patch is linearly projected to a 512-dimensional embedding, scaled by \(\sqrt{512}\), and combined with a sinusoidal positional encoding. The resulting token sequence is processed by six Transformer encoder layers, each comprising multi-head self-attention (16 heads) and a position-wise feed-forward network with residual connections and LN.

The encoder outputs are concatenated and passed through an MLP head that maps the sequence representation to a convolutional feature tensor of shape \((B,32,16,16)\). This tensor is then upsampled by the shared decoder (Fig.~\ref{fig:overview}(f)) to obtain the reconstructed image.

\subsection{GCN-based model}

The GCN-based architecture (Fig.~\ref{fig:overview}(h)) reconstructs a normalized \(128\times128\) image by treating contiguous measurement blocks as node features on a learned graph, applying stacked graph-convolutional layers, and decoding the aggregated representation via convolutional upsampling. Specifically, \(\mathbf{s}\) is partitioned into 32 graph nodes; each node corresponds to an equal-length feature block, which is linearly projected to a 512-dimensional latent representation.

Graph connectivity is defined by a trainable symmetric adjacency matrix (including self-loops) that is row-normalized and shared across the batch. We apply eight graph-convolutional layers, each implemented by a linear feature transform followed by neighborhood aggregation, BN, and ReLU; dropout (\(p=0.2\)) is used in all but the final layer. The per-node outputs are concatenated and mapped by an MLP to a tensor of shape \((B,32,16,16)\), which is subsequently upsampled by the shared decoder (Fig.~\ref{fig:overview}(f)) to yield the reconstructed image.

\subsection{Recurrent encoders (RNN/LSTM/GRU)}

The RNN-based architecture (Fig.~\ref{fig:overview}(i)) reconstructs a normalized \(128\times128\) image by interpreting \(\mathbf{s}\) as a temporal sequence, encoding the sequence with a bidirectional multi-layer RNN, and decoding the aggregated representation. Concretely, \(\mathbf{s}\) is segmented into a sequence of at most 32 equal-length steps; each step is linearly projected to a 256-dimensional embedding. The embedded sequence is processed by a three-layer bidirectional RNN (hidden-state size of 256 per direction, inter-layer dropout of \(p=0.2\)). The per-step outputs are concatenated and mapped by an MLP to a convolutional feature tensor of shape \((B,32,16,16)\), which is then upsampled by the shared decoder (Fig.~\ref{fig:overview}(f)) to generate the reconstructed image.

The LSTM-based architecture (Fig.~\ref{fig:overview}(k)) is identical to the RNN pipeline, with the recurrent cell replaced by a three-layer bidirectional LSTM of the same embedding width, hidden size, dropout, and shared decoder. The GRU-based architecture (Fig.~\ref{fig:overview}(j)) likewise replaces the recurrent cell with a three-layer bidirectional GRU of the same configuration. All three recurrent models share the same embedding width, hidden size, dropout, and the shared convolutional decoder (Fig.~\ref{fig:overview}(f)); they differ only in the recurrent cell.

\subsection{Segmentation training protocol}
\label{sec:segprotocol}

The segmentation experiments of the main text (the STSF variants, SPIFS, and the task-adapted baselines) are optimized independently of the reconstruction probe above. All learnable models are trained with AdamW (weight decay $1\times10^{-4}$) under a cosine-annealed learning-rate schedule (initial learning rate $5\times10^{-4}$); epochs and mini-batch size are set per dataset (Carvana: 100 epochs, batch~16; MNIST: 60 epochs, batch~32; WBC: 180 epochs, batch~6), and the TPLS phase fractions of main-text Eq.~(6) are $(r_{1},r_{2})=(0.3,0.3)$ for Carvana and MNIST and $(0.6,0.3)$ for WBC. During training only, images are augmented with random horizontal flips ($p=0.5$), rotations within $\pm15^{\circ}$, brightness/contrast jitter ($\pm0.2$), random crops whose side lengths span $85$--$100\%$ of the frame, and additive image-domain Gaussian noise ($\sigma=0.01$); masks receive only the geometric transforms, and validation and test images are unaugmented.

For WBC, every reported paper result and public checkpoint uses the deterministic \texttt{paper-legacy-v1} release protocol: 231 training, 58 validation, and 60 test samples. The release pipeline's generated split manifest is disjoint by sample identity, and an independent content-hash audit finds no duplicate image or mask content across the three splits. This protocol reproduces the paper and checkpoints; the repository's collision-safe full-data mode is provided only for future experiments and is not linked to the reported numbers.

The SPIFS baseline is retrained end-to-end from scratch under these settings and the shared fixed acquisition, without its original two-stage pretraining protocol. The re-implementation accordingly omits the learned (convolutional-encoder) illumination patterns of the original work, uses the same Dice+CE segmentation loss as the other methods (BCE for the binary datasets, multi-class cross-entropy for WBC) rather than the original mean-squared-error objective, and matches the decoder seed to $1\times32\times32$ ($=1024$), as in the original.

\section{Additional main-text figure}

This section collects the training-dynamics diagnostic referenced from the main text (Fig.~\ref{fig:S_training}). Per-architecture specifications for the eight probe encoders are given in Sec.~\ref{sec:archspec}.

\begin{figure*}[t]
\centering\includegraphics[width=0.86\linewidth]{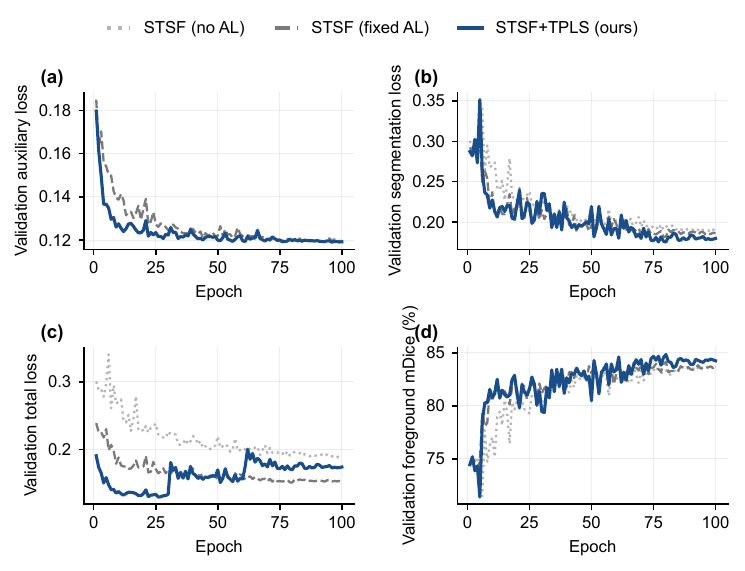}
\caption{Validation curves on the Carvana dataset at a 3.13\% sampling rate for the three STSF loss-scheduling variants: STSF (no AL), STSF (fixed AL), and the full STSF+TPLS with the proposed three-phase task-prioritized loss schedule.
(a) Auxiliary reconstruction loss (STSF (no AL) applies zero auxiliary weight and therefore has no curve in this panel).
(b) Segmentation loss.
(c) Total loss.
(d) Validation foreground mDice (\%).
On this dataset, STSF+TPLS attains lower segmentation loss and higher validation foreground mDice than the STSF (fixed AL) and STSF (no AL) variants. Panel (c) plots each variant's own training objective; because the three variants weight their loss terms differently, total-loss values are not comparable across variants. The curve in (d) is the foreground metric reported throughout the paper; model selection itself uses the class-averaged validation mDice (a convergence diagnostic).}
\label{fig:S_training}
\end{figure*}

\section{Why measurement noise reverses the ranking}
\label{sec:reversal}

The operating-regime analysis of the main text (Sec.~IV-D) reports that image-free inference overtakes the strongest reconstruct-then-segment (task-adapted, TA) baseline once measurement noise is added, despite TA winning the noiseless limit. This section shows that the TA collapse is \emph{distribution shift}, not a conditioning effect of the inverse, and derives why it is sharpest on the sparse-foreground MNIST target.

\subsection{The reconstruction adjoint is a well-conditioned isometry}

The classical inverse shared by the reconstruct-then-segment baselines is the Hadamard adjoint. Let \(\Phi\in\{\pm1\}^{M\times N}\) be the \(M\) retained rows of the \(N\times N\) Sylvester--Hadamard matrix (\(M=512\), \(N=16{,}384\)). Its rows are mutually orthogonal with squared norm \(N\), so
\begin{equation}
\Phi\Phi^{\top}=N\,\mathbf{I}_{M}.
\end{equation}
All \(M\) singular values of \(\Phi\) therefore equal \(\sqrt{N}\): the operator is perfectly conditioned on its row space (condition number \(1\)), unlike inverse operators with decaying spectra, whose small singular values amplify noise. The reconstruction is the adjoint \(\hat{\mathbf{x}}=\Phi^{\top}\mathbf{s}/N\) (zero-padding the length-\(M\) measurement vector to length \(N\) and applying the full Hadamard transform; the \(1/N\) puts the result in scene units and is cosmetic downstream, since the per-image min--max normalization is scale-invariant). For a noisy measurement \(\mathbf{s}=\Phi\mathbf{x}+\mathbf{n}\) with \(\mathbf{n}\sim\mathcal{N}(0,\sigma^{2}\mathbf{I}_{M})\),
\begin{equation}
\hat{\mathbf{x}}=\mathrm{P}_{\mathrm{row}}\,\mathbf{x}+\Phi^{\top}\mathbf{n}/N,
\end{equation}
the row-space projection of the scene ($\mathrm{P}_{\mathrm{row}}{=}\Phi^{\!\top}\Phi/N$) plus a noise term of per-pixel variance \(\sum_{k=1}^{M}\Phi_{kj}^{2}\,\sigma^{2}/N^{2}=M\sigma^{2}/N^{2}\) (each \(\Phi_{kj}^{2}=1\)). Both terms pass through the same operator \(\Phi^{\top}/N\) (a scaled isometry), so the adjoint neither amplifies nor attenuates the measured-mode signal-to-noise ratio: per measured mode, the reconstruction is no noisier than the measurements. The collapse is therefore not caused by ill-conditioning-induced signal-to-noise amplification in the Hadamard adjoint.

\subsection{The failure is distribution shift, compounded by normalization and overlap geometry}

Three coupled factors explain the collapse of the clean-trained reconstruction route.
(i) \emph{Out-of-distribution input.} At $3.13\%$ the natural-order adjoint already renders even the \emph{noiseless} scene as a low-contrast, vertically-banded image (Fig.~\ref{fig:speckle20}, top): the $M{=}512$ retained rows exercise only four of the $128$ vertical Hadamard modes, so undersampling alone, with no noise, produces the stripe/weave, making reconstruction a fragile inference intermediary under this severely undersampled protocol. On top of this the adjoint spreads the \(M\) \emph{noise} coefficients over all \(N\) pixels as a \emph{random} full-frame speckle (Fig.~\ref{fig:speckle20}, bottom).

\begin{figure}[t]
\centering\includegraphics[width=0.62\linewidth]{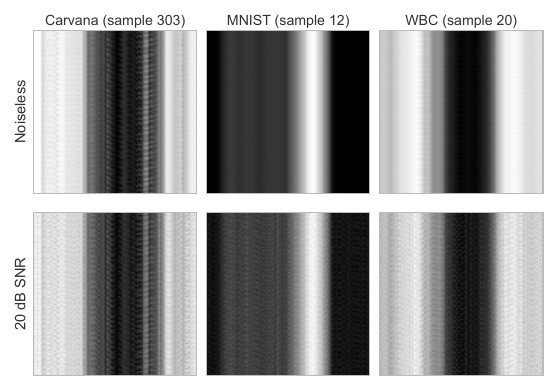}
\caption{What the reconstruct-then-segment head actually reads, noiseless versus at a 20~dB measurement SNR, for the same held-out test specimens as main-text Fig.~5 (samples 303, 12, and 20). \emph{Top (noiseless):} even without noise, the natural-order Hadamard adjoint at $3.13\%$ renders the scene as a low-contrast, vertically-banded image (the $M{=}512$ retained rows exercise only four of the $128$ vertical Hadamard modes, so undersampling alone produces the stripe/weave); the clean-trained head learns to segment through this \emph{deterministic} degradation and wins the clean limit. \emph{Bottom (20~dB):} the adjoint additionally spreads the measurement noise into a \emph{random} full-frame speckle the clean-trained head has never seen, the out-of-distribution input behind its collapse and the input on which the off-the-shelf head fails outright.}
\label{fig:speckle20}
\end{figure} The U-Net++ head is trained on the clean (banded) reconstructions and learns to segment through their \emph{deterministic} pattern, which is why it wins the clean limit; but it never sees the \emph{random} speckle that noise adds, so at test time it reads an out-of-distribution image. Learned image reconstruction is known to generate plausible false structure off-distribution \cite{recon_halluc}; the dominant signature here is instead a loss of predicted foreground, the head starving rather than imposing false structure (main-text Sec.~IV-E).
(ii) \emph{Normalization enlarges the relative task-input perturbation.} The two pipelines normalize at different points, and this placement, not the well-conditioned adjoint, sets the perturbation the segmenter actually sees. Write the normalized relative task-input perturbation as \(\delta\), the per-coordinate RMS perturbation of its input divided by that input's min--max normalization range. Image-free decoding min--max normalizes the raw measurement vector, whose range is anchored by the large DC coefficient \(s_{0}\) (the summed scene intensity); the AC-referenced noise is only a small \emph{absolute} fraction of that range, so \(\delta_{\mathrm{IF}}\) stays small. The task-adapted pipeline instead min--max normalizes the \emph{reconstruction}, in which the same DC arrives not as a coordinate but as a spatially \emph{uniform} plane (the all-ones Hadamard row): per-image min--max is shift-invariant, so no such normalization can anchor the range to the DC (the implementation's explicit mean removal \(\tilde{\mathbf{r}}=\hat{\mathbf{x}}-\overline{\hat{\mathbf{x}}}\) is an identity operation for the subsequent min--max, and the CS variant omits it with the same outcome). Its range \(R_{\mathrm{TA}}=\max\tilde{\mathbf{r}}-\min\tilde{\mathbf{r}}\) is therefore set by the spatial variation alone: the AC image content \emph{together with} the full-frame noise speckle, with no stable anchor. For the same injected measurement noise, the normalized perturbation scales inversely with the normalization range, so the ratio
\begin{equation}
A=\frac{\delta_{\mathrm{TA}}}{\delta_{\mathrm{IF}}}\ \propto\ \frac{R_{\mathrm{IF}}}{R_{\mathrm{TA}}}
\label{eq:amplify}
\end{equation}
is governed by the ratio of the DC-anchored range to the DC-blind image range: the range factor exceeds unity whenever the DC coefficient dominates the dynamic range (on these scenes it also overcomes the \(N/\sqrt{M}\) dimension deficit), and it grows with the DC-to-AC anchoring ratio of the measurement vector, which sets how strongly the measurement-domain range is pinned while the image-domain range is not. The amplification factorizes into two computable terms. Per coordinate, the noise reaching each pipeline is exact: in the measurement domain each of the \(M\) entries carries noise of RMS \(\sigma\); in the image domain the adjoint gives \(e=\Phi^{\top}\mathbf{n}/N\) with \(\mathbb{E}[e_j^2]=M\sigma^{2}/N^{2}\), i.e.\ per-pixel RMS \(\sigma\sqrt{M}/N\). Dividing each by its normalization range (the range treated as noise-free to first order; the end-to-end measurement below includes all effects),
\begin{equation}
A=\frac{\delta_{\mathrm{TA}}}{\delta_{\mathrm{IF}}}
=\frac{\bigl(\sigma\sqrt{M}/N\bigr)/R_{\mathrm{TA}}}{\sigma/R_{\mathrm{IF}}}
=\underbrace{\frac{\sqrt{M}}{N}}_{\text{dimension factor}}\times\underbrace{\frac{R_{\mathrm{IF}}}{R_{\mathrm{TA}}}}_{\text{range factor}},
\label{eq:amplify_split}
\end{equation}
where the dimension factor \(\sqrt{M}/N=\sqrt{512}/16384\approx1.4\times10^{-3}\) strongly \emph{favors} the reconstruction (the well-conditioned adjoint dilutes the noise energy over \(N\) pixels), so the measured ratio is carried entirely by the range factor, which must also overcome this deficit. Measured directly on the full test splits, the DC-anchored measurement range \(R_{\mathrm{IF}}\) exceeds the DC-blind image range \(R_{\mathrm{TA}}\) by roughly four orders of magnitude (\({\approx}8.9{\times}10^{3}\) on MNIST, \({\approx}4.2{\times}10^{4}\) on Carvana, and \({\approx}4.3{\times}10^{4}\) on WBC; the DC-to-AC ratio \(s_{0}/\mathrm{std}(s_{\mathrm{AC}})\) alone contributes \({\approx}21\), \({\approx}228\), and \({\approx}188\)), for a first-order prediction \(A\approx12\)--\(59\) from Eq.~\eqref{eq:amplify_split}. Measured end to end on the real Hadamard operator with real test images (full test splits, three noise seeds, the deployed pipeline's 8-bit round-trip included), the normalized relative task-input perturbation of the reconstruction exceeds that of the image-free bucket at a 20~dB SNR by \(A{\approx}20\times\) on MNIST, \({\approx}53\times\) on Carvana, and \({\approx}70\times\) on WBC (\(3.5\)--\(6.7\%\) versus \({\leq}0.4\%\); the ADMM variant reads \(4.0\)--\(8.1\%\), \(A{\approx}24\)--\(84\times\)), within a factor of \(1.7\) of the first-order prediction on every dataset for the HSI pipeline, the residual being the noise's own contribution to the min--max anchors that the first-order treatment freezes.
(iii) \emph{Overlap geometry.} Foreground mIoU divides intersection by union. Where the foreground is sparse (MNIST strokes fill a small fraction of the frame), speckle false-positives inflate the union denominator far more than for a frame-filling object (Carvana), driving the ratio toward zero. This is why MNIST shows the steepest TA fall and the largest reversal margin (\(+21.7\)~pp at 20~dB), while Carvana degrades gently.

Direct image-free decoding reads the measurement sequence without forming this full-frame image and remains much closer to its clean training distribution across the tested noise grid. This is consistent with the graceful-degradation motivation of task-oriented inference and joint source--channel coding \cite{task_oriented_comm, deep_jscc}, but the evidence here is specific to the tested SPS protocol. We test the out-of-distribution account~(i) directly below and quantify the normalization effect~(ii) both by Eq.~\eqref{eq:amplify} and by direct measurement; the overlap-geometry contribution~(iii) we establish by argument.

\subsection{Demonstrating the mechanism: noise-augmented retraining}

The distribution-shift account makes a falsifiable prediction: if the task-adapted head fails under test-time noise only because it was trained on clean reconstructions, then retraining it on \emph{noisy} reconstructions should restore accuracy at that noise level. We generate the HSI reconstructions with the same AC-calibrated noise injected into the bucket before reconstruction, and retrain the U-Net++ head on them: a fixed-\(20\)~dB head and a mixed-SNR head (per-image SNR drawn uniformly from \(10\)--\(40\)~dB). Main-text Table~III reports foreground mIoU on MNIST across the noise sweep.

The intervention confirms the mechanism and sharpens the claim. The clean-trained head collapses from \(57.9\) at \(40\)~dB to \(36.0\) at \(20\)~dB; the fixed-\(20\)~dB head fully recovers \emph{at} \(20\)~dB (\(57.8\), essentially the clean-trained head's clean-limit accuracy) but falls off its trained point (\(49.1\) at the unseen \(15\)~dB, where the clean-trained head reads \(13.2\); \(33.5\) at \(10\)~dB), and the mixed-SNR head, robust across its training range, still degrades at the hard end (\(49.9\) at \(10\)~dB). Image-free STSF+TPLS, trained only on clean measurements, stays flat across the tested grid (\(56.3\!\rightarrow\!56.1\)) without noise-specific retraining. The task-adapted collapse is thus out-of-distribution error, not an information limit: reconstruction \emph{can} be made noise-robust near its training calibration, whereas the fixed and mixed-SNR probes degrade away from that support. The operating-regime reversal reflects this difference under the tested calibration and noise range.

\section{Robustness under physically faithful shot (Poisson) noise}
\label{sec:poisson}

The main-text noise study injects additive Gaussian noise into the bucket signal, calibrated on its AC-coupled component. A single-pixel detector is, however, photon-limited: its dominant corruption is shot (Poisson) noise, whose variance grows with the collected flux. We therefore repeat the operating-regime comparison under a physically faithful shot-noise model and ask whether the reversal survives.

\subsection{Shot-noise model}

Each $\pm1$ Hadamard measurement is realized on the binary DMD as a complementary pattern pair, $s_i = c^+_i - c^-_i$, where $c^\pm_i$ are the photon counts under the two complementary patterns and $c^+_i + c^-_i = T$, the total scene flux (the DC measurement, $\mathrm{raw}[0]$). For a nonnegative scene, $|s_i|\le T$; hence $c^\pm_i = (T \pm s_i)/2 \ge 0$. We scale the total flux to a photon budget $K$ per measurement ($\kappa = K/T$), draw shot-noisy counts $\tilde c^\pm_i \sim \mathrm{Poisson}(\kappa\, c^\pm_i)$, and form $\tilde s_i = (\tilde c^+_i - \tilde c^-_i)/\kappa$. Since $\mathrm{Var}[\tilde s_i]=(\kappa c^+_i+\kappa c^-_i)/\kappa^{2}=T^{2}/K$, the shot floor on each measurement is $T/\sqrt{K}$, set by the shared \emph{absolute} budget $K$ rather than by per-scene renormalization as in the additive model. We sweep $K$ from $10^8$ (bright) to $10^2$ (photon-starved) photons per measurement. At $K{=}10^8$ every method recovers its clean-limit accuracy (Table~\ref{tab:poisson}: reconstruction $87.3/57.6/53.0$ vs.\ the noiseless TA-HSI $87.9/57.5/53.1$ on Carvana/MNIST/WBC), which validates the calibration.

\subsection{The reversal survives}

Table~\ref{tab:poisson} reports foreground mIoU across the photon-budget sweep. The two families respond oppositely, as under additive noise. The reconstruction path collapses under photon starvation. The adjoint spreads the shot noise of every measurement into a full-frame speckle that the clean-trained head cannot parse, driving foreground mIoU toward zero (reaching ${\approx}0$ below ${\sim}10^{4}$ photons on MNIST and already below ${\sim}10^{6}$ on WBC). Image-free inference degrades far more gracefully and stays above the reconstruction path throughout the photon-limited regime, so it overtakes reconstruction at a physical, dataset-dependent photon budget bracketed by the measured decades: between $10^{5}$ and $10^{6}$ photons per measurement on MNIST, between $10^{6}$ and $10^{7}$ on Carvana, and between $10^{7}$ and $10^{8}$ on WBC (Table~\ref{tab:poisson}). The regime reversal established under additive noise thus holds under the \emph{dominant physical noise} of single-pixel sensing, and can be stated as an operating point in photons.

Two qualifications. First, the near-invariance of image-free inference in the moderate regime is partly a pipeline property: the per-sample min--max normalization of the measurement vector is anchored by the large DC coefficient (total flux), which keeps the per-coefficient shot noise a small fraction of the network input (cf.\ Sec.~\ref{sec:reversal}), so the robustness is of the measurement-domain pipeline, not an inherent network noise-immunity. Image-free accuracy nonetheless falls once the budget is starved enough (already at $K{=}10^{5}$ on Carvana, and on all three datasets by $K \le 10^{3}$), confirming the noise is not merely normalized away. Second, under \emph{extreme} starvation the ranking \emph{among} image-free lifts is dataset-dependent: on the frame-filling Carvana target, whose large flux carries the most absolute shot noise into the high-frequency coefficients, the simpler SPIFS projection stays higher than the STSF cross-attention lift (Table~\ref{tab:poisson}). This is the imprinting signature of main-text Sec.~IV-E rather than superior sensing; SPIFS is flat because it largely ignores its input (on MNIST its accuracy moves by at most $0.1$~pp from $10^{8}$ down to $10^{2}$ photons), so its shot robustness is degenerate. Both image-free methods remain above the collapsed reconstruction path, and the reversal is a statement about image-free versus reconstruction, which holds on all three datasets. As in the main text this sweep is in simulation; we do not couple $K$ to the noise-equivalent power of a specific detector, and the physical bench remains proof of concept.

\begin{table}[t]
\centering
\caption{Operating-regime comparison under physically faithful shot (Poisson) noise on MNIST/Carvana/WBC for the two image-free methods and the strongest reconstruct-then-segment baseline (foreground mIoU, mean over three noise seeds; single training seed; bold = best method at that budget). Photon budget $K$ = mean photons per measurement (shot floor $T/\sqrt{K}$). At $K{=}10^8$ both arms recover the clean limit; as photons are starved, the reconstruction path collapses toward $0$ while image-free inference degrades gracefully and overtakes it; the reversal survives shot noise, crossing over at a dataset-dependent photon budget. SPIFS's flatness across budgets reflects its input-insensitivity (imprinting, main-text Sec.~IV-E), not superior sensing.}
\label{tab:poisson}
\begin{tabular}{llccccccc}
\hline
Dataset & Method & $10^8$ & $10^7$ & $10^6$ & $10^5$ & $10^4$ & $10^3$ & $10^2$ \\
\hline
\multirow{3}{*}{MNIST} & STSF+TPLS (image-free) & 56.5 & 56.5 & 56.5 & \textbf{56.5} & \textbf{56.5} & \textbf{55.6} & \textbf{52.6} \\
 & SPIFS (image-free) & 42.6 & 42.6 & 42.6 & 42.6 & 42.6 & 42.6 & 42.7 \\
 & TA (reconstruction) & \textbf{57.6} & \textbf{57.8} & \textbf{58.0} & 46.5 & 11.2 & 0.1 & 0.0 \\
\hline
\multirow{3}{*}{Carvana} & STSF+TPLS (image-free) & 74.6 & 74.6 & \textbf{74.3} & 54.2 & 40.1 & 37.0 & 36.2 \\
 & SPIFS (image-free) & 69.7 & 69.7 & 69.6 & \textbf{69.0} & \textbf{63.3} & \textbf{51.7} & \textbf{46.1} \\
 & TA (reconstruction) & \textbf{87.3} & \textbf{78.6} & 58.0 & 47.3 & 35.9 & 31.6 & 30.8 \\
\hline
\multirow{3}{*}{WBC} & STSF+TPLS (image-free) & 52.7 & \textbf{52.7} & \textbf{52.7} & \textbf{52.4} & \textbf{49.7} & 39.4 & 36.6 \\
 & SPIFS (image-free) & 49.1 & 49.1 & 49.1 & 48.9 & 47.7 & \textbf{44.4} & \textbf{41.9} \\
 & TA (reconstruction) & \textbf{53.0} & 51.0 & 38.7 & 0.0 & 0.0 & 0.0 & 0.0 \\
\hline
\end{tabular}
\end{table}

\section{Acquisition-order ablation}
\label{sec:perm}

A permutation of the measurement sequence, applied jointly to the measurements and the \emph{known} pattern schedule, loses no information, so a temporal encoder cannot derive its advantage from the natural Hadamard order carrying information, and the central architectural motivation must not rest on the order being physically meaningful. We test this directly: we retrain STSF+TPLS on MNIST under a fixed, known random permutation of the Hadamard pattern schedule (identical information content, architecture, and training seed), and compare against the natural (Sylvester) order. Foreground mIoU is $56.7$ under the permuted schedule versus $56.5$ under the natural order, a $0.1$~pp difference at a single matched seed ($56.68$ vs.\ $56.54$ before rounding), well within the $0.47$~pp cross-seed spread measured for STSF+TPLS on MNIST (main-text Sec.~IV-E), i.e.\ practically indistinguishable. The natural acquisition order therefore carries no information advantage. Accordingly, we motivate the temporal encoder not on any special ordering but on \emph{parameter-efficient sequence processing at a matched budget}: the capacity-controlled encoder comparison of the main text shows that a fully connected map, which reads the same measurement vector with no sequential inductive bias, is consistently outperformed by the recurrent gated encoders at an equal parameter count, an architectural effect rather than an order effect.

\section{Gradient interference under the TPLS schedule}
\label{sec:gradconflict}

The main text (Sec.~III-B) motivates TPLS by gradient interference between the auxiliary-reconstruction and segmentation objectives. We measure this directly. Along the \emph{actual} training trajectory (MNIST, $M{=}512$, the deployed TPLS configuration with the schedule compressed to $9$ epochs at the same phase fractions, $40$ steps per epoch, batch $32$, seed $42$; five identical instrumented repeats), each step computes the two gradients separately, $\nabla_{\Theta}\mathcal{L}_{\mathrm{Seg}}$ and $\nabla_{\Theta}\mathcal{L}_{\mathrm{AL}}$, records their cosine similarity and magnitude ratio, and then takes the ordinary optimizer step on the TPLS-weighted sum. Table~\ref{tab:gradconflict} reports the resulting per-phase statistics.

\begin{table}[h]
\centering
\caption{Per-phase gradient statistics, mean over five identical instrumented runs (seed~42, epochs~1--9, per-step means within each run); brackets give the run-to-run range across the five repeats. The measurement is sensitive to run-to-run nondeterminism (nondeterministic CUDA backward atomics, amplified over the early trajectory), so the values characterise the trend rather than fixing an exact point---but the trend is stable across every repeat. The gradients are near-orthogonal in \emph{direction} throughout (small $|\cos|$, roughly half the steps mildly negative), so hard directional conflict is rare; the interference is one of \emph{magnitude}: as the segmentation loss converges, the raw ratio $\lVert\nabla\mathcal{L}_{\mathrm{AL}}\rVert/\lVert\nabla\mathcal{L}_{\mathrm{Seg}}\rVert$ rises ${\sim}6\times$ across training and crosses unity. Under a fixed equal weight the auxiliary term would therefore \emph{dominate} the update late in training, whereas the TPLS weights $(\alpha,\beta)$ keep the effective auxiliary-to-task pull $\alpha\lVert\nabla\mathcal{L}_{\mathrm{AL}}\rVert/\beta\lVert\nabla\mathcal{L}_{\mathrm{Seg}}\rVert$ bounded and decreasing (from ${\approx}2.4$ to ${\approx}0.2$), and conversely amplify the initially ${\approx}4\times$ weaker auxiliary gradient in phase~1. GMS is the gradient-magnitude similarity $2\lVert g_{\mathrm{AL}}\rVert\lVert g_{\mathrm{Seg}}\rVert/(\lVert g_{\mathrm{AL}}\rVert^{2}+\lVert g_{\mathrm{Seg}}\rVert^{2})\in(0,1]$, averaged per step (not the GMS of the mean ratio): it is far from $1$ exactly when one task's gradient dwarfs the other's, quantifying that the imbalance TPLS corrects is a magnitude, not a direction, phenomenon. We deliberately do not probe loss-surface curvature; the trajectory statistics above are first-order and make no curvature claim.}
\label{tab:gradconflict}
\begin{tabular}{lccccc}
\hline
Phase ($\alpha_{\mathrm{aux}},\beta_{\mathrm{seg}}$) & $\overline{\cos}$ & $\Pr[\cos<0]$ & $\overline{\lVert g_{\mathrm{AL}}\rVert/\lVert g_{\mathrm{Seg}}\rVert}$ & GMS & effective pull \\
\hline
1 $(0.9, 0.1)$ & $+0.005$ & $0.46$ & $0.26$ {\scriptsize[0.23,\,0.31]} & $0.47$ {\scriptsize[0.42,\,0.52]} & $2.37$ {\scriptsize[2.07,\,2.79]} \\
2 $(0.5, 0.5)$ & $-0.000$ & $0.47$ & $1.04$ {\scriptsize[0.82,\,1.37]} & $0.84$ {\scriptsize[0.79,\,0.88]} & $1.04$ {\scriptsize[0.82,\,1.37]} \\
3 $(0.1, 0.9)$ & $+0.016$ & $0.36$ & $1.65$ {\scriptsize[1.28,\,2.00]} & $0.84$ {\scriptsize[0.79,\,0.88]} & $0.18$ {\scriptsize[0.14,\,0.22]} \\
\hline
\end{tabular}
\end{table}

This measurement refines the main-text reading: TPLS is best understood not as resolving \emph{directional} gradient conflict (there is little to resolve) but as a magnitude rebalancing schedule that matches the auxiliary prior's influence to the optimization stage, strong while the task gradient is large and noisy, waning as the task converges and the raw auxiliary gradient would otherwise take over. Statistics are means over five repeats of this configuration on one dataset; the full per-step records are archived.

\section{Clean-condition qualitative comparison}
\label{sec:cleanqual}

Figure~\ref{fig:s14clean} complements the noisy-operating-point comparison of the main text (main-text Fig.~5) with the same three-dataset qualitative comparison under \emph{noiseless} acquisition, and shows the full method ladder including the off-the-shelf reconstruct-then-segment baseline (main-text Fig.~5 displays it at the noisy operating point). The two figures share the WBC specimen (test index 20), so the clean and 20~dB conditions can be compared directly on the same cell.

\begin{figure*}[t]
\centering\includegraphics[width=0.78\linewidth]{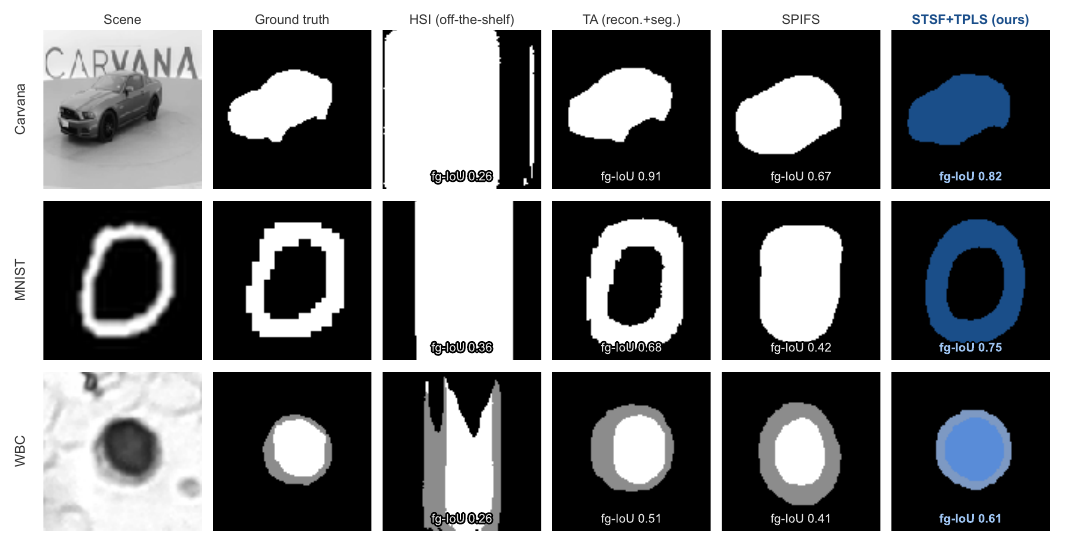}
\caption{The clean operating point, qualitatively: noiseless segmentation on the three test sets at a $3.13\%$ sampling rate (held-out test samples 34, 2, and 20; per-sample foreground IoU on each mask). Columns: scene (context only, not available to any single-pixel method), ground truth, then the full reconstruct-then-segment ladder: the \emph{off-the-shelf} U-Net++ (trained on fully-sampled images, so the $3.13\%$ reconstruction is already a train--test sampling-rate mismatch) collapses on its stripe artifacts, while the \emph{task-adapted} (TA) retraining of the same segmenter turns the identical reconstruction into crisp, boundary-faithful masks, sharpest on Carvana ($0.91$); image-free SPIFS (rounded blob) and STSF+TPLS follow. Clean acquisition is the regime where reconstruction, once task-adapted, is strongest; main-text Fig.~5 shows the ordering reverse at a 20~dB measurement SNR.}
\label{fig:s14clean}
\end{figure*}

\section{Qualitative results at ultra-low sampling rates}
\label{sec:lowrate}

The main text reports that image-free accuracy holds across a $32\times$ sampling-rate span (main-text Fig.~4(a--c)). Figure~\ref{fig:lowrate} makes this axis visible on the same specimens as the main-text qualitative comparison, with one model per rate and method (single seed; SPIFS and TA-HSI rows are included for comparison). For STSF+TPLS the masks remain object-shaped and correctly placed at $0.39\%$ sampling ($64$ measurements for a $128^2$ scene, $256\times$ fewer samples than pixels): the car silhouette stays complete (foreground IoU $0.73$ versus $0.83$ at $3.13\%$), the digit remains legible as a ``7'' ($0.56$ versus $0.60$), and the cell keeps its nucleus--cytoplasm topology ($0.52$ versus $0.61$); the degradation is a graceful coarsening, not a collapse, consistent with the flat quantitative plateau. The added rows show the other two regimes of Table~\ref{tab:rates} on the same specimens: TA-HSI sharpens as the rate rises, its clean-limit advantage visible at $3.13\%$ on Carvana and MNIST, while SPIFS is unstable across rates. Classical reconstruction, by contrast, is already severely degraded as an image at $3.13\%$ under the same fixed natural-order patterns, although a task-adapted head can exploit its deterministic clean structure (Fig.~\ref{fig:speckle20}, top).

\begin{figure*}[t]
\centering\includegraphics[width=0.68\linewidth]{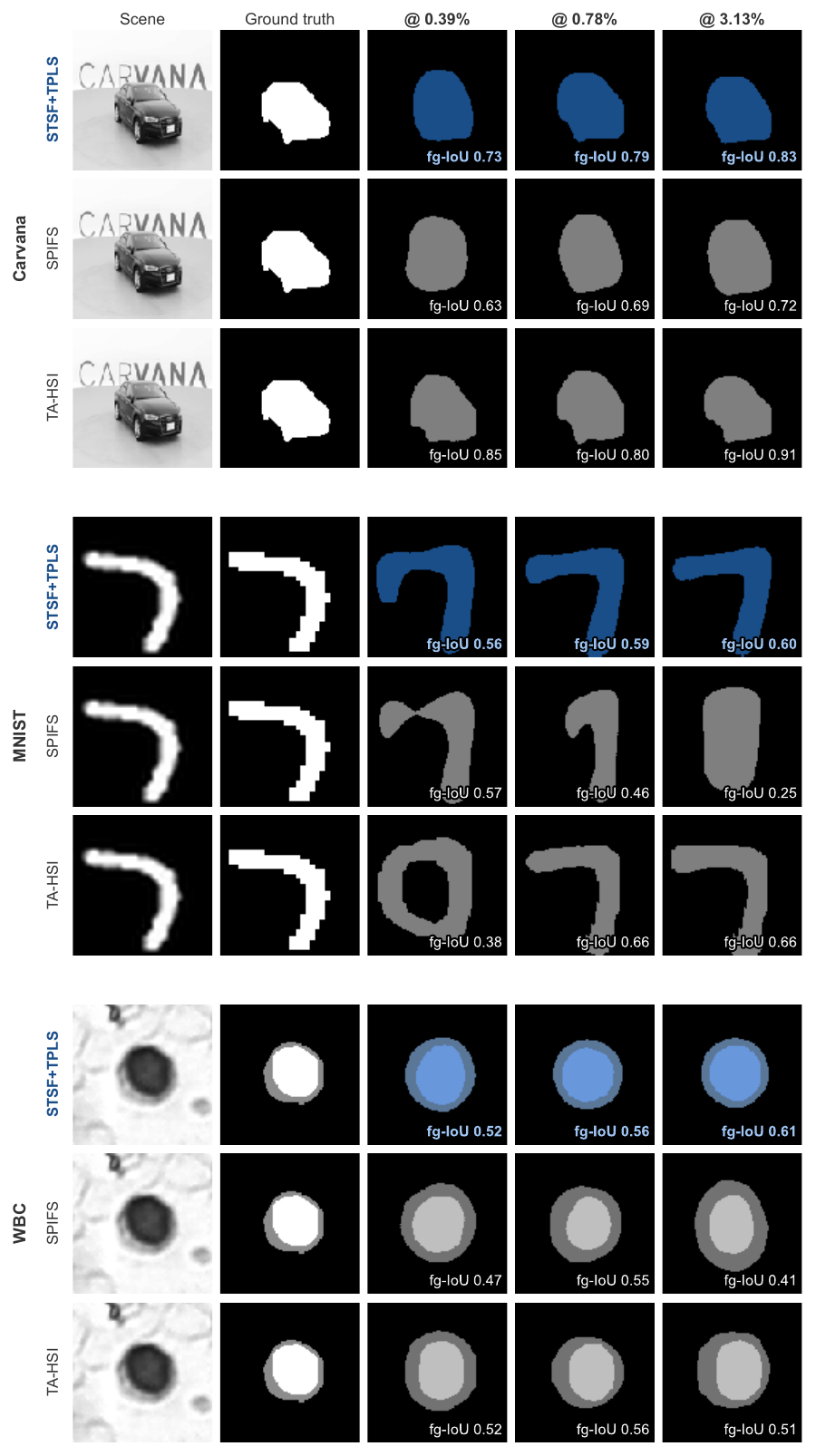}
\caption{The sampling-rate axis, qualitatively: STSF+TPLS, SPIFS, and TA-HSI masks at $0.39\%$, $0.78\%$, and $3.13\%$ sampling on the specimens of the main-text qualitative figure (one model per rate and method, single seed; per-mask foreground IoU annotated). STSF+TPLS coarsens gracefully and stays object-shaped down to $0.39\%$; TA-HSI generally sharpens with rate (Table~\ref{tab:rates}), though not monotonically on every specimen shown; SPIFS is unstable across rates on the MNIST and WBC specimens.}
\label{fig:lowrate}
\end{figure*}

Table~\ref{tab:rates} tabulates the full sweep behind main-text Fig.~4(a--c): the image-free arms and the task-adapted (TA) baselines at all six rates on all three datasets.

\begin{table}[ht]
\centering
\caption{Foreground mIoU across the sampling-rate sweep (single training seed, one model per rate; the tabulated counterpart of main-text Fig.~4(a--c)). The task-adapted (TA) baselines (HSI and CS) are evaluated at all six rates on all three datasets, one model trained per rate. On Carvana and MNIST their accuracy rises steeply with rate, tracking the reconstruction quality the segmenter consumes; on the three-class WBC target the TA curves are flatter and non-monotonic, staying within a $49$--$57$ band across the whole $32\times$ span. Bold: best method per dataset and rate.}
\label{tab:rates}
\begin{tabular}{ll cccccc}
\hline
Dataset & Method & 0.39\% & 0.78\% & 1.56\% & 3.13\% & 6.25\% & 12.5\% \\
\hline
\multirow{4}{*}{Carvana} & STSF+TPLS & 68.7 & 73.1 & 73.7 & 74.6 & 76.3 & 74.7 \\
 & SPIFS & 64.3 & 69.5 & 64.5 & 69.7 & 65.5 & 61.9 \\
 & TA-HSI & \textbf{81.9} & 81.6 & \textbf{83.6} & \textbf{87.9} & 90.2 & \textbf{95.5} \\
 & TA-CS & 77.8 & \textbf{81.7} & 83.2 & 86.6 & \textbf{90.4} & 95.0 \\
\hline
\multirow{4}{*}{MNIST} & STSF+TPLS & \textbf{54.3} & 57.0 & 56.7 & 56.5 & 56.7 & 57.3 \\
 & SPIFS & 53.4 & 49.9 & 47.6 & 42.6 & 51.0 & 46.6 \\
 & TA-HSI & 53.4 & \textbf{57.6} & 57.3 & \textbf{57.5} & 64.9 & \textbf{75.8} \\
 & TA-CS & 53.5 & 57.4 & \textbf{57.5} & 57.4 & \textbf{66.6} & 75.0 \\
\hline
\multirow{4}{*}{WBC} & STSF+TPLS & 49.1 & 53.5 & 53.2 & 52.7 & 53.2 & 53.8 \\
 & SPIFS & 47.9 & 51.1 & 48.6 & 49.1 & 48.3 & 47.2 \\
 & TA-HSI & 49.1 & 53.8 & \textbf{55.4} & 53.1 & 54.1 & \textbf{56.9} \\
 & TA-CS & \textbf{49.8} & \textbf{55.0} & 53.3 & \textbf{54.3} & \textbf{54.9} & 55.9 \\
\hline
\end{tabular}
\end{table}

\section{Per-example failure-mode gallery}
\label{sec:gallery}

Figure~\ref{fig:gallery} is the qualitative counterpart of the failure signatures quantified in main-text Fig.~6: five held-out MNIST scenes (rows) segmented by each method (columns) at a 20~dB measurement SNR (noise seed 1, the same AC calibration as the quantitative panels and main-text Fig.~5), selected from a scan of all 2,000 test scenes as the sharpest exemplars of each mode. The fixed-physics reconstruction path (TA-HSI) collapses: the stroke thins to a placed but ink-starved residue retaining only $17$--$20\%$ of the ground-truth foreground (rows 1--3), or shatters into fragments (rows 4--5, a ``0'' into arcs, a ``3'' into blobs). The learned-static SPIFS returns the same blob regardless of input (imprinting), and the content-adaptive STSF+TPLS coarsens while staying correctly placed, preserving the loop of the ``0'' and both loops of the ``8'' (coarsening).

The column headers name each method's \emph{dominant} tendency. These are dominant tendencies, not an exclusive taxonomy: a minority of the reconstruction path's failures ($5.9\pm5.5\%$ across training seeds; main-text Sec.~IV-E) are instead confident, full-ink wrong shapes, and the adaptive lift's rare failures are over-full rather than absent. The gallery shows the typical mode of each region of the lift spectrum, not a label that holds on every scene.


\section{Temporal-encoder selection results}
\label{sec:encoder_probe}

The main text uses reconstruction only as an offline, task-agnostic proxy for selecting the temporal encoder. Table~\ref{tab:encoder_probe} provides the complete probe results; it does not describe the deployed segmentation path, and its single-run protocol supports an architecture choice rather than a seed-level performance claim.

\begin{table}[h]
\centering
\caption{Reconstruction probe used for temporal-encoder selection: PSNR (dB, $\uparrow$) and SSIM ($\uparrow$) on MNIST at three sampling rates, and on Fashion-MNIST and CelebA at a 3.13\% sampling rate. The parameter column lists model size at 3.13/6.25/12.5\% (the 3.13\% count applies to Fashion-MNIST and CelebA). Every cell is a single training run. The same checkpoint-selection rule is applied to all architectures, so the comparison is internally controlled, but the absolute values are optimistic and the ranking has no seed uncertainty. Best value per column in bold.}
\label{tab:encoder_probe}
\resizebox{\textwidth}{!}{%
\begin{tabular}{l c cc cc cc|cc cc}
\hline
\multirow{2}{*}{Models} & \multirow{2}{*}{Params (M)} &
\multicolumn{2}{c}{MNIST 3.13\%} &
\multicolumn{2}{c}{MNIST 6.25\%} &
\multicolumn{2}{c|}{MNIST 12.5\%} &
\multicolumn{2}{c}{FMNIST 3.13\%} &
\multicolumn{2}{c}{CelebA 3.13\%} \\
\cline{3-12}
 & & PSNR & SSIM & PSNR & SSIM & PSNR & SSIM & PSNR & SSIM & PSNR & SSIM \\
\hline
Physics-based & N/A & 9.02 & 0.431 & 9.17 & 0.430 & 9.92 & 0.450 & 10.06 & 0.378 & 10.35 & 0.278 \\
FC & 136.4/138.5/142.7 & 13.88 & 0.699 & 13.61 & 0.691 & 13.52 & 0.690 & 15.94 & 0.611 & 15.08 & 0.437 \\
CNN & 141.1/145.0/152.6 & 13.37 & 0.697 & 12.97 & 0.680 & 12.90 & 0.677 & 15.40 & 0.606 & 14.88 & 0.434 \\
U-Net & 132.5/140.9/157.7 & 13.29 & 0.686 & 13.20 & 0.680 & 13.03 & 0.688 & 15.25 & 0.604 & 14.64 & 0.431 \\
GCN & 136.4/136.4/136.4 & 13.15 & 0.679 & 11.67 & 0.661 & 12.17 & 0.405 & 14.02 & 0.545 & 14.27 & 0.413 \\
Transformer & 146.9/146.9/146.9 & 14.50 & 0.722 & 14.55 & 0.723 & 16.25 & 0.769 & 15.75 & 0.610 & 14.67 & 0.422 \\
RNN & 135.3/135.3/135.3 & 13.06 & 0.675 & 12.57 & 0.659 & 12.90 & 0.666 & 15.72 & 0.601 & 13.16 & 0.367 \\
LSTM & 138.5/138.5/138.5 & 16.05 & 0.765 & 14.36 & 0.717 & 16.16 & 0.771 & 18.25 & 0.684 & 15.67 & \textbf{0.454} \\
GRU & 137.4/137.4/137.4 & \textbf{16.32} & \textbf{0.768} & \textbf{15.74} & \textbf{0.759} & \textbf{17.13} & \textbf{0.797} & \textbf{18.37} & \textbf{0.686} & \textbf{15.69} & 0.453 \\
\hline
\end{tabular}%
}
\end{table}

\FloatBarrier
\begin{figure}[htbp]
\centering
\includegraphics[width=0.53\linewidth]{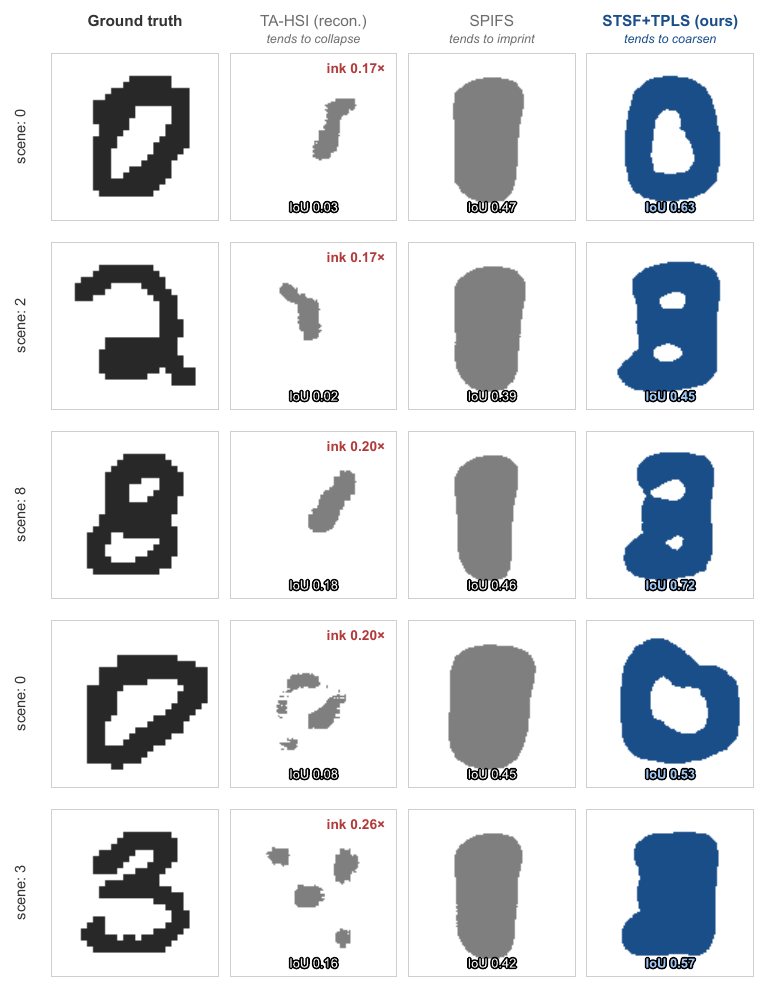}
\caption{Per-example failure-mode gallery (five held-out MNIST scenes $\times$ three methods, 20~dB measurement SNR). Column headers give each method's \emph{dominant} tendency (collapse / imprinting / coarsening); these are dominant tendencies across the lift spectrum, not exclusive per-scene labels (see text). Per-mask foreground IoU is annotated; the TA column additionally reports the retained foreground fraction (ink, TA foreground divided by ground-truth foreground). The quantitative signatures of these modes---output self-similarity, catastrophic-failure rate, and mean occupancy---are in main-text Fig.~6.}
\label{fig:gallery}
\end{figure}
\vspace{1.5em}

\else
  \includepdf[pages=-]{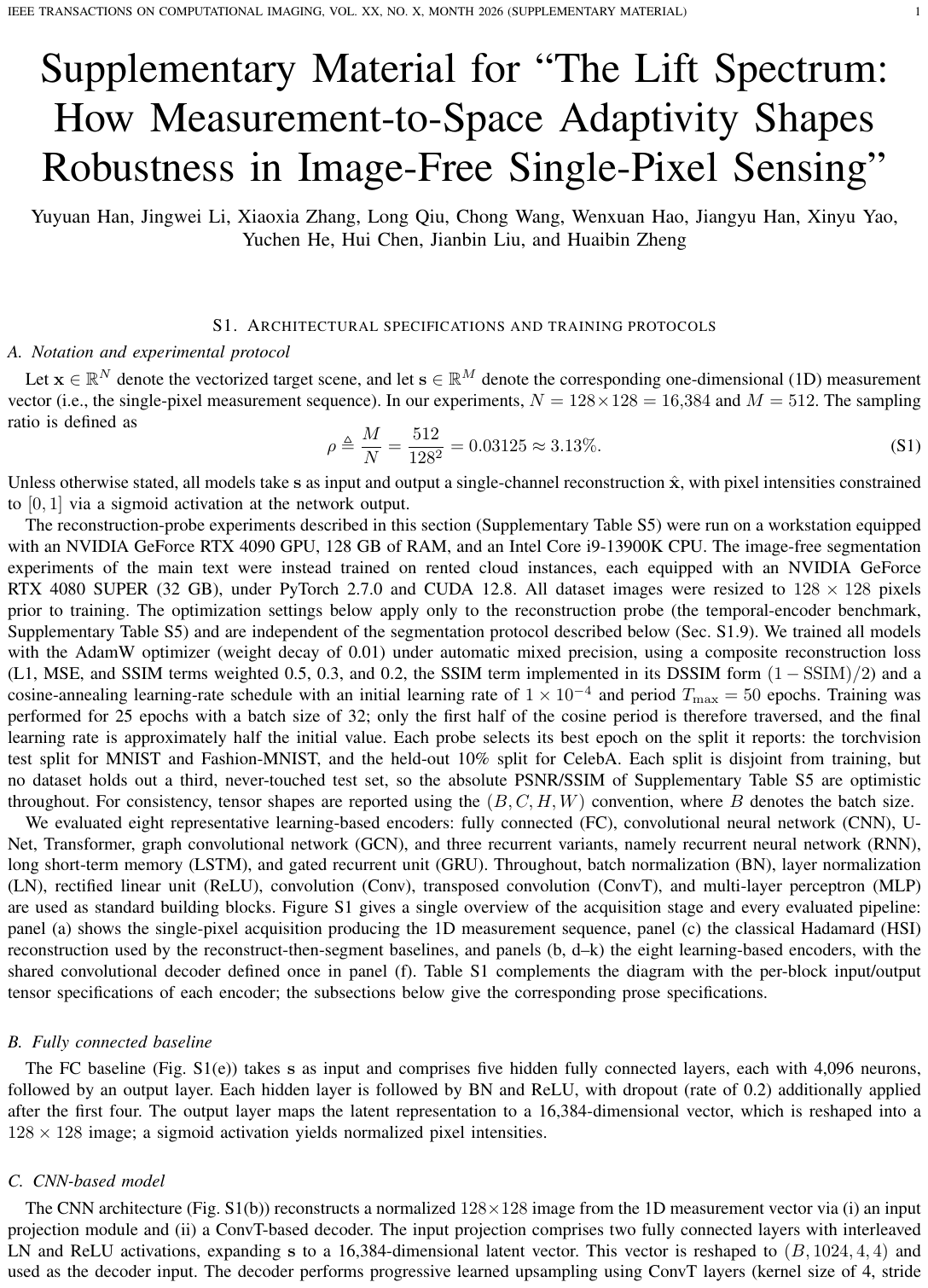}
\fi

\end{document}